\documentclass[nofootinbib,superscriptaddress]{revtex4}
\usepackage{amsfonts,amssymb,amsthm}

\usepackage{amsmath,amssymb,bbm}
\usepackage{amsfonts}
\usepackage{graphicx}

\newcommand{\C}{{\mathbb C}}
\newcommand{\N}{{\mathbb N}}
\newcommand{\R}{{\mathbb R}}
\newcommand{\Z}{{\mathbb Z}}

\newcommand{\cE}{{\mathcal E}}
\newcommand{\cF}{{\mathcal F}}

\newcommand{\cK}{{\mathcal K}}

\newcommand{\cH}{{\mathcal H}}

\newcommand{\cP}{{\mathcal P}}

\newcommand{\cC}{{\mathcal C}}
\newcommand{\cS}{{\mathcal S}}
\newcommand{\cV}{{\mathcal V}}
\newcommand{\cX}{{\mathcal X}}

\newcommand{\XX}{{\bf X}}
\newcommand{\kamone}{{\frac{1}{\kappa}}}

\newcommand{\stpm}{{$*$-product}}
\newcommand{\demi}{{\frac{1}{2}}}
\newcommand{\one}{\mbox{$1 \hspace{-1.0mm}  {\bf l}$}}

\newcommand{\SU}{\mathrm{SU}}
\newcommand{\DSU}{\mathrm{DSU}}
\newcommand{\Spin}{\mathrm{Spin}}

\newcommand{\SO}{\mathrm{SO}}
\newcommand{\U}{\mathrm{U}}

\newcommand{\su}{\mathfrak{su}}

\newcommand{\be}{\begin{equation}}
\newcommand{\ee}{\end{equation}}
\newcommand{\beq}{\begin{eqnarray}}
\newcommand{\eeq}{\end{eqnarray}}
\newcommand{\bes}{\begin{eqnarray}}
\newcommand{\ees}{\end{eqnarray}}
\newcommand{\bea}{\begin{eqnarray}}
\newcommand{\eea}{\end{eqnarray}}
\newcommand{\nn}{\nonumber}

\newcommand{\mat} [2] {\left ( \begin{array}{#1}#2\end{array} \right ) }

\newcommand{\la}{\langle}
\newcommand{\ra}{\rangle}

\newcommand{\w}{\wedge}

\newcommand{\tr}{{\mathrm Tr}}
\newcommand{\f}{\frac}

\def\pp{{\partial}}
\def\arr{\rightarrow}
\def\mone{^{-1}}
\def\dr{\rightarrow}

\def\ka{\kappa}
\def\eps{\epsilon}

\newcommand{\id}{\mathbb{I}}

\def\tE{\tilde{E}}
\def\hf{\widehat{f}}
\def\hphi{\widehat{\phi}}
\def\hpsi{\widehat{\psi}}

\def\vX{\vec{X}}

\def\vp{\vec{p}}
\def\vP{\vec{P}}
\def\vsigma{\vec{\sigma}}
\def\hu{\hat{u}}
\def\tw{\widetilde{w}}
\def\bz{\bar{z}}
\def\deltas{\delta^\star}

\def\tdelta{\widetilde\delta}

\begin{document}

\title{Spinors and Voros star-product for Group Field Theory: First Contact}

\author{{\bf Maite Dupuis}}\email{maite.dupuis@ens-lyon.fr}
\affiliation{Laboratoire de Physique, ENS Lyon, CNRS-UMR 5672, 46 All\'ee d'Italie, Lyon 69007, France}
\affiliation{School of Physics, The University of Sydney, Sydney, New South Wales 2006, Australia}

\author{{\bf Florian Girelli}}\email{girelli@physics.usyd.edu.au}
\affiliation{School of Physics, The University of Sydney, Sydney, New South Wales 2006, Australia}

\author{{\bf Etera R. Livine}}\email{etera.livine@ens-lyon.fr}
\affiliation{Laboratoire de Physique, ENS Lyon, CNRS-UMR 5672, 46 All\'ee d'Italie, Lyon 69007, France}
\affiliation{Perimeter Institute, 31 Caroline St N, Waterloo ON, Canada N2L 2Y5}

\date{\today}

\begin{abstract}

In the context of non-commutative geometries, we
develop a group Fourier transform for the Lie group $\SU(2)$. Our method is based on the Schwinger representation of the Lie algebra $\su(2)$ in terms of spinors. It allows us to prove that the non-commutative $\R^3$ space dual to the $\SU(2)$ group is in fact of the Moyal-type and endowed with the Voros star-product when expressed in the spinor variables. Finally, from the perspective of quantum gravity, we discuss the application of these new tools to group field theories for spinfoam models and their interpretation as non-commutative field theories with quantum-deformed symmetries.

\end{abstract}

\maketitle



\section*{Introduction}


Spinfoam models provide us with a framework for regularized path integral  for quantizing gravity (for a review, see e.g. \cite{review-spinfoam}). They define transition amplitudes for quantum states of geometry and can be considered as the covariant definition of Loop Quantum Gravity.
The first spinfoam model can be seen retrospectively as the the Ponzano-Regge model for 3d Euclidian gravity (with no cosmological constant). The path integral is defined on discretized 3d manifold and the resulting quantum gravity partition function is essentially defined in terms of the $6j$-symbols of the recoupling theory of $\SU(2)$ representations.
Since then the spinfoam framework has been much developed and generalized to the four-dimensional case and refined in order to account for a Lorentzian signature and a cosmological constant and to incorporate matter fields \cite{new-models}.

 \smallskip

In the early 90's, Boulatov  showed \cite{boulatov} that the spinfoam amplitudes of the Ponzano-Regge model  could be obtained as Feynman diagram amplitudes of a non-local (quantum) field theory  defined over a Lie group manifold, in this case $\SU(2)^{\times 3}$. It was later shown that all spinfoam models can be reformulated in such terms and generated from a group field theory (GFT) \cite{gft}.
The introduction of GFTs to generate spinfoam amplitudes was an important technical development since it allows to sum in a controlled way over topologies and now the GFTs are considered as the proper non-perturbative definition of spinfoam models.
Moreover the GFT framework allows to discuss the issue of spinfoam continuum limit and semi-classical limit in terms of renormalization \cite{review-gft}.  A field theory formulation is then a perfect framework to address  the typical divergencies one meets in the spinfoam approach, for instance in the infra-red regime \cite{gft-divergences}.

 \smallskip

Until recently, the usual point of view on group field theories was to consider the Lie group manifold as the configuration space and perform a Fourier transform using the Peter-Weyl theorem in order to obtain the spinfoam amplitudes in terms of representations and $nj$-symbols for the relevant group.
In the recent years, non-commutative techniques entered  the game  thanks to a new type of  Fourier transform \cite{gftnc1,gftnc2,gftsym1,gftsym2,gftmatter2}.  The general mathematical formalism behind this generalized Fourier transform was mostly developed by Majid \cite{majid} and it was   rediscovered later on in the context of 3d spinfoam while coupling particles to the Ponzano-Regge model \cite{fourier0, fourier1}.
%
From this perspective, the Lie group manifold is now interpreted as the momentum space and a group Fourier allows us to go from the group to the dual configuration space, which is then a non-commutative space of the Lie algebra type.

\smallskip

In the context of Boulatov's GFT, this generalized Fourier transform formalism has been used to construct the relevant non-commutative $\R^3$ configuration space dual to the group manifold.
As a consequence, the GFT can be understood as a non-commutative field theory. However this construction was only done for the group $\SO(3)=\SU(2)/\Z_2$ whereas to recover the full Ponzano-Regge spinfoam model, one should use  $\SU(2)$. The Fourier transform for a momentum space given by $\SU(2)$ turns out to be more difficult to construct in order to have  a one-to-one map from momentum space to configuration space.
In an earlier work \cite{fourier3}, the authors analyzed very carefully the details  of the group Fourier transform  pinpointing that the typical choice of plane-waves on $\SU(2)$leads to a two-to-one map.  In \cite{fourier2}, the authors nevertheless constructed a  one-to-one Fourier transform for $\SU(2)$ from a 4d point of view. The first result we will present  here is the construction of a well-defined one-to-one Fourier transform for $\SU(2)$, using a 3d realization.

\smallskip

This non-commutative perspective for spinfoams allows to connect the diffeomorphism symmetry (in 3d) of spinfoam amplitudes to  quantum group symmetries of the field theory \cite{gftsym2}.  One can expect that these symmetries will be useful in order to discuss the question of renormalization of the field theory by putting constraints on the renormalization scheme and allowed counter-terms. Another interesting strength for the spinfoam non-commutative perspective is that it allows to connect  spinfoams to deformed special relativity (DSR) which is a candidate phenomenological model  to encode  effective quantum gravity corrections to matter kinematics and dynamics in the semi-classical regime \cite{gftmatter2}.
Unfortunately, quantum field theories based on non-commutative spaces of the Lie algebra type\footnote{We have in mind the ones which are not constructed by a simple twist \cite{tolstoi}, such as $\su(2)$ or $\ka$-Minkowski. }  are very poorly understood at this time. For example, there is no integral form for the $\SO(3)$ and $\SU(2)$ star products; there is no fermionic or Yang-Mills theories defined yet;  the divergence structure of such quantum field theories is barely known, in general.

In fact, the non-commutative field theories which have attracted the most attention are the Moyal non-commutative field theory, ie a non-commutative space of the type $[x_\mu,x_\nu]=\theta_{\mu\nu}.$ In this context, Yang-Mills theories have been introduced and very detailed analysis of  quantum field theories have been performed \cite{review-moyal}. Our second result consists in showing  that a GFT based on $\SU(2)$ can also be  seen as some sort of Moyal field theory, more exactly a non-commutative field using a star product of the Voros type (see \cite{voros1,voros2} for different perspectives on the Voros quantum field theory). We expect  that this will open new doors to address the issue of renormalization in quantum gravity.
The key-idea in deriving this result and obtaining  the Voros non-commutative product is to consider the Jordan-Schwinger representation for $\su(2)$. This representation consists in introducing a pair of harmonic oscillators, or a spinor $|z\ra\in\C^2$, to describe the Lie algebra $\su(2)$.
This  spinor formalism for spin networks and spinfoam models,  is inspired by the $\U(N)$ formalism for intertwiners \cite{un1,un2,un3,un4} and twisted geometries for loop quantum gravity \cite{twisted1,twisted2}. It has been further developed in \cite{spinor}.

\smallskip

In the first section,  we recall the construction of the $\SO(3)$ Fourier transform and the issue with generalizing to $\SU(2)$.
%
In the second section, we recall the spinor construction and introduce the plane-waves and star product defined in terms of the spinor variables. We show that the Fourier transform based on this "spinor plane-wave"  is well-defined   for $\SU(2)$.
%
In the third section, we discuss the implications  of the spinor representation. In particular we show how we can recover the 4d bicovariant differential calculus naturally. We also show that the star product constructed using the spinor plane-waves actually co\"incides with the Voros star product.
%
%
In the fourth section, we apply the results of the previous sections to the GFT context, focusing in particular on the Boulatov model. Explicitly we present the new shape of the closure constraint using the spinor variables, and make explicit Boulatov action in terms of the spinor variables. We conclude by discussing the quantum group symmetries of the model.
%
We have added two appendices. In the first one,   we recall the notion of coherent intertwinners which is relevant to defining the non-commutative delta function in configuration space. In the second one, we discuss  the different choices of plane-waves one can make using the spinor variables.

%
%
%

%
%
%
%
%


In the following we will always work in units $\hbar=c=1$ and  $\kappa$ is a mass scale, usually taken to be the Planck mass in the context of quantum gravity (phenomenology).

\section{Star-product for $\SO(3)$ and Fourier transform: an overview} \label{so3}

In the context of the matter coupling to the Ponzano-Regge spinfoam model for 3d Euclidean quantum gravity, it's been understood that particles and fields behave as in a non-commutative flat geometry \cite{fourier0,fourier1}. Indeed with the particle momenta now living on the Lie group manifold $\SU(2)$, which is curved, the natural space-time coordinates defined as dual to the momentum coordinates are naturally non-commutative. This is the same mechanism as happening in deformed or doubly special relativity (DSR) in four space-time dimensions when deforming the Poincar\'e symmetry in order to accommodate a universal Planck length (e.g. \cite{DSR}).

To make this relation between momentum living in $\SU(2)$ and non-commutative 3d coordinate space, a group Fourier transform between $\SU(2)$ and $\R^3_\ka$ was first introduced in \cite{fourier0,fourier1} and further developed in \cite{fourier2,fourier3}. This allowed to describe the propagation of matter coupled to the 3d quantum geometry in terms of actual space-time coordinates. To be more precise, the original group Fourier transform introduced in the context of spinfoam models in \cite{fourier0,fourier1} maps functions on $\SO(3)\sim \SU(2)/\Z_2$ to functions on $\R^3$. Later on in \cite{fourier2}, this group Fourier transform was refined to truly go between $\SU(2)$ and $\R^3$, but we will here first focus on the original map between $\SO(3)$ and $\R^3$, which is currently used to provide spinfoam models and group field theories with a space-time interpretation.

In the framework of the Ponzano-Regge spinfoam model, the natural candidate for a Fourier transform between functions on $\SU(2)$ and functions on $\R^3$ is:
\be
\label{attempt1}
\hf(\vX)=\int_{\SU(2)} dg\, f(g)\,e^{\f\ka{2}\tr gX}
\qquad\textrm{with}\quad
X=\vX\cdot\vsigma \in\su(2),
\ee
where $\sigma_i$ are the  Pauli matrices (normalized such that $\sigma_i^2=\id$ for $i=1,2,3$ and  $\sigma_i\sigma_j= \delta_{ij}\id + i\epsilon_{ijk} \sigma_k$.). Using the standard parametrization of $\SU(2)$ group elements as $2\times 2$ matrices,
\be
g=\cos\theta\id+i\sin\theta\hu\cdot\vsigma
\qquad\textrm{with}\quad
\theta\in[-\pi,\pi],\quad
\hu\in\cS^2,
\ee
we easily evaluate the exponent:
\be
\f\ka{2}\tr gX = \,i\,\vp\cdot\vX,\qquad
\textrm{with}\quad
\vp=\ka\,\f1{2i}\tr g\vsigma,
\quad
g=\eps\sqrt{1-\f{p^2}{\ka^2}}+i\f\vp\ka\cdot\vsigma\,,
\ee
where the momentum is bounded in norm, $|p|\le \ka$, and $\eps=\pm$ registers the sign of $\cos\theta$. In this context, it is natural to introduce a $\star$-product between the plane waves $e_g(X)\equiv e^{\f\ka{2}\tr gX}$ which keeps track of  the group multiplication on $\SU(2)$:
\be
(e_{g_1}\star e_{g_2})(X)\,=\,e_{g_1g_2}(X).
\ee

The problem with this proposal is that the group Fourier transform defined by \eqref{attempt1} has a non-trivial kernel:
$$
f(-g^{-1})\,=\,-f(g)\quad\Rightarrow\quad
\hf(\vX)=0.
$$
This comes because the momentum conjugated to the coordinates $\vX$ is the 3-vector $\vp$ defined as the projection of the group element $g$ onto the Pauli matrices, but that the map $g\arr\vp$ is not a bijection but is two-to-one. This can be seen directly when trying to recovering the $\delta$-distribution on $\SU(2)$ by the inverse Fourier transform:
\be
\int d^3\vX\,e^{\f\ka2\tr gX}\propto
\delta(g)+\delta(-g)
\propto\delta_{\SO(3)}(g),
\ee
as first pointed out in \cite{pr1}. Thus it seems more natural to define a group Fourier transform from $\SO(3)$ to $\R^3$ if using these plane waves $e_g(X)= e^{\f\ka{2}\tr gX}$. Thus following \cite{fourier0,fourier1}, we modify our definition of the group Fourier transform \eqref{attempt1} and define instead:
\be
\label{attempt2}
\hf(\vX)=\int_{\SU(2)} dg\, f(g)\,e^{\f\ka{2}\tr |g|X}\,,
\ee
where we define the absolute value of a group element as $|g|=g$ if $\cos\theta\ge 0$ else $|g|=-g$ if $\cos\theta\le 0$. The plane wave exponent is now:
$$
\f\ka2\tr |g|X =\,i\eps\vp\cdot\vX\,.
$$
This absolute value satisfies the obvious identities:
$$
|g|=|-g|,\qquad
|g_1g_2|
\,=\,
|\,|g_1|\,|g_2|\,|\,.
$$
And in terms of $p$-momentum, it reads as:
\be
g(\vp,\eps)=\eps\sqrt{1-\f{p^2}{\ka^2}}+i\f\vp\ka\cdot\vsigma,
\qquad
|g|=\sqrt{1-\f{p^2}{\ka^2}}+i\eps\f\vp\ka\cdot\vsigma=g(\eps\vp,+).
\ee
It is then natural to restrict ourselves to functions on $\SO(3)$, i.e. even functions on $\SU(2)$ satisfying $f(g)=f(-g)$. Thus defining $f(\vp)=f(\vp,+)=f(-\vp,-)$, the $\SO(3)$ group Fourier transform reads:
\be
\hf(\vX)=\int_{|p|<\ka} \f{d^3\vp}{\pi^2\ka^3\sqrt{1-\f{p^2}{\ka^2}}}\,
f(\vp)\,e^{i\vp\cdot\vX},
\ee
where the non-trivial measure in $\vp$ reflects the normalized Haar measure on $\SO(3)=\SU(2)/\Z_2$. It is then natural to introduce a $\star_s$-product inherited from the group multiplication on $\SO(3)$:
\be
e^{\f\ka2\tr |g_1|X}
\,\star_s\,
e^{\f\ka2\tr |g_2|X}
\,=\,
e^{\f\ka2\tr |\,|g_1|\,|g_2|\,|X}
\,=\,
e^{\f\ka2\tr |g_1g_2|X}\,.
\ee
This $\star_s$-product can be translated into a modified addition on momenta in the $\vp$ variables:
$$
e^{i\vp_1\cdot\vX}
\,\star_s\,
e^{i\vp_2\cdot\vX}
=
e^{i(\vp_1\oplus\vp_2)\cdot\vX},
$$
with the following deformed addition law:
\be\label{sumso3}
\vp_1\oplus\vp_2
=
\eps_{12}\left(\sqrt{1-\f{p_2^2}{\ka^2}}\vp_1+\sqrt{1-\f{p_1^2}{\ka^2}}\vp_2
-\f1\ka\vp_1\w\vp_2
\right),
\ee
where $\eps_{12}$ is the sign of $\sqrt{1-\f{p_1^2}{\ka^2}}\sqrt{1-\f{p_2^2}{\ka^2}}-\f1{\ka^2}\vp_1\cdot\vp_2$. This sign flip is a necessary subtlety of this group Fourier transform for $\SO(3)$.
Expanding this formula for small momentum, we can compute the commutator between coordinates:
\be
[X_i,X_j]_{\star_s}=\f2\ka i\eps_{ijk}X_k,
\ee
which shows explicitly the non-commutativity structure of space-time.

\medskip

Furthermore, using the inverse Fourier transform of the $\delta$-distribution,
\be
\f1{(2\pi)^3}\int d^3\vX\,e^{\f\ka2\tr gX}
=\f1{(2\pi)^3}\int d^3\vX\,e^{\f\ka2\tr |g|X}
=\delta(g)+\delta(-g)
=2\delta_{\SO(3)}(g),
\ee
we can use the $\star_s$ product to write the inverse Fourier transform for general even functions on $\SO(3)$:
\be
f(g)=\f1{2(2\pi)^3}\int d^3\vX\, \hf(\vX)\,\star_s\, e^{\f\ka2\tr |g^{-1}|X}
=
\f1{2(2\pi)^3}\int d^3\vX\, \hf(\vX)\,\star_s\, e^{-\f\ka2\tr |g|X}.
\ee
We can also use the explicit parametrization in terms of $\vp$ to give an explicit formula for the inverse:
\be
f(\vp)=\sqrt{1-\f{p^2}{\ka^2}}\int \f{\ka^3}{(2\pi)^3}{d^3\vX}\, \hf(\vX)e^{-i\vp\cdot\vX},
\ee
for functions $\hf(\vX)$ with a standard Fourier transform with support on momentum bounded by $\ka$ in norm.

The Fourier transform of the matrix elements and characters of $\SO(3)$ group elements can also be computed. They are expressed in terms of Bessel functions. We refer the interested reader to \cite{fourier1,fourier2,fourier3,gftnc1}.

\medskip

Finally, we would like to remind the reader that there is an ambiguity in the choice of the momentum variable on which is based the whole construction. Instead of choosing plane waves $\exp(i\vp\cdot\vX)$ in terms of the momentum $\vp\propto\tr g\vsigma$, one could choose place waves $\exp(i\vP\cdot\vX)$ based on different choice of parametrization of the group elements $g\in\SO(3)$. These leads to different Fourier transforms and star-products e.g. \cite{fourier3,Bobs}. For instance, choosing $\vP=\ka\tan\theta \hu = \ka\tr g\vsigma /\tr g$ avoids the issue of having a bounded momentum and it is still possible to define the star-product and deformed addition of momenta. Nevertheless, $\vp$ seems to be the nicest choice with respect to the differential calculus \cite{fourier2,fourier3}.

\section{Spinor plane-waves and $\star$-product for $\SU(2)$}

In this section, we will show how to use the recently developed spinorial tools for $\SU(2)$ to define new plane-waves and a group Fourier transform on the whole $\SU(2)$ group.


\subsection{Spinors and 3-vectors} \label{spinor}

The spinor formalism for spin networks and spinfoam models \cite{un1,un2,un3,un4}  is based on the simple remark that 3-vectors can be constructed as the projection of spinors on Pauli matrices and that we have the natural action of $\SU(2)$ on spinors as $2\times 2$ matrices.

More explicitly, let us start with a spinor $z\in\C^2$. This is a two-dimensional complex vectors living in the fundamental representation of $\SU(2)$. We will use the ket-bra notations:
$$
|z\ra=\mat{c}{z_0 \\ z_1},\qquad
\la z|=\mat{cc}{\bz_0 & \bz_1}.
$$
Then we consider the Hermitian matrix $|z\ra\la z|$, from which define the dimensionful vector   $\vX \in\R^3$:
\be\label{def X}
\vX\,=\,
\f1\ka\tr |z\ra\la z|\vsigma
=\kamone \la z|\vsigma|z\ra= \kamone  \overline{z}^a \vsigma _{ab} z^b,
\qquad
|z\ra\la z|=\frac{\kappa}{2}\left(
|\vX|\id+\vX\cdot\vsigma
\right),\quad \textrm{ with }\,\,|\vX|=\frac{\la z|z\ra}{\kappa}.
\ee
The vector $\vX$ entirely determines the original spinor $z$ up to a global phase $z\arr e^{i\alpha} z$. Then all $\U(1)$-invariant functions of the spinor $z$ are functions of $\vX$ and vice-versa.
The change of integration variable from $d^4z$ to a measure $d^4\mu(\vec X,\phi)$ can be easily computed. In particular, for a $\U(1)$-invariant function $f$, we can show that:
\be
\f1{\pi^2}\int d^4z\,e^{-\la z|z\ra}\,f(\vec{X}(z))
\,=\,
\f1{4\pi}\int \f{d^3\vec X}{|\vec X|}\,e^{-|\vX|}\,f(\vec X)\,.
\ee
\smallskip

It is natural to endow the space of spinors $\C^2$ with the canonical Poisson bracket $\{z_a,\bz_b\}=-i\delta_{ab}$. This induces the following brackets on the $X_i$ coordinates.
\bes
\{X_i,X_j\}&=&\,\frac{2}{\kappa}\eps_{ijk}X_k\,, \\
\{X_i,|\vec X|\}&=&0.
\ees
Thus the $X_i$'s form a $\su(2)$ algebra and actually generate the fundamental $\SU(2)$ action on spinors. $|\vec X|$ gives the (square root of the) $\su(2)$ Casimir.  At the quantum level, this simply becomes the Schwinger representation for $\su(2)$ in terms of a couple of harmonic oscillators.

The present proposal exploits this expression of a 3-vector $\vX$ in terms of a spinor $z$ and uses the fact that the action of $\SU(2)$ group elements on $z\in\C^2$ simply induces the corresponding 3d rotation on the vector $\vX$. Using the fundamental two-dimensional representation of $\SU(2)$ instead of the three-dimensional action of $\SU(2)$ on 3-vectors will avoid the problem of only representing $\SO(3)$ and will allow to define a Fourier transform for all functions on $\SU(2)$.

\subsection{ Spinor plane-waves on $\SU(2)$}

Following the previous work on the spinorial formulation of $\SU(2)$ and its representation theory \cite{holo2,spinor}, a natural candidate for the new $\SU(2)$ plane wave is:
\be
E_g(z)\,\equiv\,
e^{\la z|g|z\ra}
\,=\,
e^{\tr g\,|z\ra\la z|}.
\ee
This functional of the spinor $z$ is clearly invariant under the multiplication of the spinor by a global phase, so it can be expressed solely in terms of the 3-vector $\vX$:
\be
E_g(\vX)=\,
e^{\frac{\kappa}{2}|\vec X|\,\tr g}\,e^{\frac{\kappa}{2}\tr gX}.
\ee
Comparing to the $\SO(3)$ plane waves discussed earlier, there are two differences:
\begin{enumerate}
\item There is a new phase factor depending on the norm $|\vX|$ and on the trace of the group element. This trace $\tr g$ allows to distinguish $g$ from $-g$ and thus allows to probe the whole $\SU(2)$ group.
\item We do not need to take the absolute value of the group element and the main factor of the plane wave is $e^{\frac{\kappa}{2}\tr gX}$ and not $e^{\frac{\kappa}{2}\tr |g|X}$ as earlier.
\end{enumerate}

In terms of the $(\vp,\eps)$ parametrization, these new spinorial plane waves read:
\be
E_{g(\vp,\eps)}(\vX)=\,
e^{\eps \kappa  |\vec X|\sqrt{1-\f{p^2}{\ka^2}}}\,e^{ i \vp\cdot\vX},
\ee
with the special pre-factor depending on the norms of $\vX$ and $\vp$.

\medskip

As an element of $\cC(\SU(2))$, the plane-wave $e^{\la z|g|z\ra}$ is square integrable for the Haar measure $dg$ of $\SU(2)$. To prove this, we notice that $\overline{\la z|g|z\ra}=\la z|g^{-1}|z\ra$, and  use the $\SU(2)$ coherent states technology as well as  the Peter-Weyl theorem (cf appendix \ref{CS}).
\bes
\int dg\,|e^{\la z|g|z\ra}|^2
&=&
\int dg\,e^{\la z|g|z\ra}e^{\la z|g^{-1}|z\ra}
=
\sum_{j,k}\f1{(2j)!(2k)!}\,
\int dg\,\la j,z|g|j,z\ra\la k,z|g^{-1}|k,z\ra
=
\sum_{j\in\N/2}\f{(\la z|z\ra^2)^{2j}}{(2j)!^2(2j+1)}\ \nn\\
&=&
\f{I_1(2\la z|z\ra)}{\la z|z\ra} 
\ees
where $I_n$ is the $n$-th modified Bessel function of the first kind. Note that when $\kappa \dr \infty$, this becomes divergent, that is we recover in the classical limit plane-waves which are not square-integrable.

\medskip

The plane-wave $e^{\la z|g|z\ra}$ can also be seen as a function of  $\vec X$. We note $\cC_\star(\R^3)$ the set of functions generated by the $\vec X$.
$\star$ denotes the  star-product between the elements in $\cC_\star(\R^3)$, which generalizes the notion of point-wise product.
We can give a precise definition of the $\star$-product by defining it on the spinorial plane-waves:  it reflects the group multiplication on $\SU(2)$.
\be\label{def star prod 1}
(E_{g_1}\star E_{g_2})(\vec X)
\,\equiv\,
E_{g_1g_2}(\vec X)\,\,\Longleftrightarrow \, \,
e^{\la z|g_1|z\ra}\star e^{\la z|g_2|z\ra}
\,=\,
e^{\la z|g_1g_2|z\ra}\,.
\ee
The  identity for this $\star$-product is  $\one _\star\equiv\,E_1(\vec X)= e^{\la z|z\ra}= e^{\ka |\vec X|}$. This is not the usual identity, given by the constant function equal to 1. Indeed, we check that:
\bes
(E_{1}\star E_{g})(\vec X) = (E_{g}\star E_{1})(\vec X)
\,\equiv\,
E_{g}(\vec X)\,\,\Longleftrightarrow \, \,
e^{\la z|z\ra}\star e^{\la z|g|z\ra}=e^{\la z|g|z\ra}\star e^{\la z|z\ra}
\,=\,
e^{\la z|g|z\ra}\,.
\ees
Note that the feature of having a non-trivial identity in  configuration space was  already 
present in \cite{fourier2}, where another Fourier transform on $\SU(2)$ (as opposed to $\SO(3)$) was introduced. In \cite{fourier2}, the authors deal with a 4d Fourier transform with a 4d momentum space defined as $\R^+\times \SU(2)$. In our scheme, we have not introduced an extra momentum dimension. As a consequence, we shall see in section \ref{kernel}, that this feature of having non-trivial identity can   be easily avoided.
Nevertheless, we can also see our construction from a 4d perspective. Indeed the algebra $\cC_\star(\R^3)$ can be seen as the subalgebra of $\cC_\star(\C^2)=\cC_\star(\R^4)$ which is generated by  functions of the spinor $z$ invariant under global phase transformations $z\dr e^{i \alpha}z$ (or equivalently the functions which $\star$-commute with $|\vec X|$ as we shall see in sections \ref{proprietes} and \ref{voros}).

\subsection{Fourier transform on $\SU(2)$ and its inverse}\label{su2}

We use the plane-wave $E_g(z)=e^{\la z|g|z\ra}$ based on the spinor variable $z$  to define a new  Fourier transform $\cF$ between $\cC(\SU(2))$ and $\cC_\star(\R^3)$
\bes
\cF: \cC(\SU(2))&\dr &\cC_\star(\R^3),\nn\\
f&\mapsto & \hf(z) = \int_{\SU(2)} dg\,f(g)\,E_g(z)
\quad\textrm{or equivalently}\quad
\hf(\vX)=\int dg\,f(g)\,e^{\f\ka2|\vec X|\,\tr g}\,e^{\f\ka2\tr gX}\,.
\ees
Since the plane-waves $E_g(z)$ are square-integrable with respect to the Haar measure $dg$ on $\SU(2)$, this Fourier transform is a well-defined map\footnotemark in the sense that $\hf(z)$ is finite for all $z\in\C^2$ provided that $f$ is in $L^2(\SU(2))$.
\footnotetext{
This can be checked directly using the Cauchy-Schwarz inequality in order to derive an explicit bound on the norm of $\hf(z)$:
$$
|\hf(z)|^2
\,=\,
\left|\int_{\SU(2)} dg\,f(g)\,e^{\la z|g|z\ra}\right|^2
\,\le\,
\left(\int dg\,|f(g)|^2\right)\,\left(\int dg\,|e^{\la z|g|z\ra}|^2\right)
\,\le\,
\f{I_1(2\la z|z\ra)}{\la z|z\ra}\,\int |f|^2
\,<\,+\infty\,.
$$
}

The $\star$-product between $\hphi,\,\hpsi\in \cC_\star(\R^3)$ is as usual the Fourier transform of the  convolution product  between the  functions $\phi,\,\psi\in \cC_\star(\SU(2))$:
\bes\label{convolution star}
\widehat{(\phi\circ \psi)}(z)&=& \int [dg]^2 dh\,  \phi(g_1)\psi(g_2) \delta(g_1g_2 h\mone) E_h(z)\nn\\
&=& \int [dg]^2 \,  \phi(g_1)\psi(g_2)  E_{g_1g_2 } (z)= \int [dg]^2 \,  \phi(g_1)\psi(g_2)  (E_{g_1}\star E_{g_2 }) (z)\nn\\
&=& (\hphi \star \hpsi) (z).
\ees

\medskip

To prove that we are really dealing with $\SU(2)$ and not $\SO(3)$, we  can compute the Fourier transform of the matrix elements of the $\SU(2)$ group elements, which form a basis of $L^2$-functions over $\SU(2)$ by the Peter-Weyl theorem.
Following the approach of \cite{un4,holo1,holo2,spinor}, we use the overcomplete basis of $\SU(2)$ coherent states labeled by a spin $j\in\N/2$ indicating the $\SU(2)$ irreducible representation and by a spinor $z\in\C^2$ defining the state. The reader can find more details on these coherent states and the corresponding decomposition of the identity in appendix \ref{CS}. Here, we will simply use that the matrix elements\footnotemark{ } of $g\mone\in\SU(2)$ on these coherent states have a simple expression:
\be
\la j,w|g\mone |j,\tw\ra = \la w|g\mone |\tw\ra^{2j}\,.
\ee
Their Fourier transform is straightforward to compute:
\bes
\hf_{w,\tw}^{(j)}(z)
&\,=\,&
\int dg\,e^{\la z|g|z\ra}\,\la j,w|g\mone |j,\tw\ra
\,=\,
\sum_k\f1{(2k)!}\int dg\,\la k,z|g|k,z\ra\,\la j,w|g\mone |j,\tw\ra
\,=\,
\f1{(2j+1)!}\,\la w|z\ra^{2j}\la z|\tw\ra^{2j}\,\nn\\
&=& \f1{(2j+1)!}\,\la w|z\ra^{2j}\la z|\tw\ra^{2j} e^{-\la z| z\ra}\, \one_\star.
\ees
\footnotetext{
A similar calculation can be done for $\la j,w|g |j,\tw\ra$ by using the fact that $g$ is unitary and taking it complex conjugate:
$$
\la j,w|g |j,\tw\ra=\overline{\la j,\tw|g\mone |j,w\ra}
=\la j,\overline{\tw}|\overline{g\mone} |j,\overline{w}\ra
=\la j,\overline{\tw}|\eps\mone{g\mone}\eps|j,\overline{w}\ra
=\la j,\eps\,\overline{\tw}|{g\mone}|j,\eps\,\overline{w}\ra,
\qquad
\textrm{with}\quad
\eps=\mat{cc}{0& -1\\ 1 &0}\,.
$$
.}
Note that we have made apparent the non-trivial identity, which brings the extra factor $e^{-\la z| z\ra}$.
The matrix elements are therefore maps to a linear combinations of polynomials of the type $\la w|z\ra^{2j}\la z|\tw\ra^{2j} \la z|z\ra^{2k}$ which are homogenous of identical degree in $|z\ra$ and $\la z|$.  They can be expressed in terms of the 3-vector, $\forall j,k\in \N/2$
\be
\la w|z\ra^{2j}\la z|\tw\ra^{2j} \, \la z|z\ra^{2k}
\,=\,
\left(\f\ka{2}\right)^{2j}
\left(| \vX|\la w|\tw\ra+  \vX\cdot\la w|\vsigma|\tw\ra\right)^{2j}\, (\ka|\vec X|)^{2k}\, .
\ee
Polynoms with terms of identical degree in $|z\ra$ and $\la z|$ are clearly a basis of all $\U(1)$-invariant polynomials of $z$ which generate $\cC_\star (\R^3)$.
This direct calculation of the Fourier transform of the matrix element functionals on $\SU(2)$ ensures that our Fourier transform does not have any non-trivial kernel as the $\SO(3)$ group Fourier transform  reviewed in section \ref{so3}. Moreover, it shows that every function in $L^2(\SU(2))$ has a finite well-defined Fourier transform since they can be decomposed onto the matrix elements.

\medskip

We can now recover the $\delta$-distribution on $\SU(2)$ as a superposition of our new spinorial plane waves. The fastest way to proceed is to use the $\SU(2)$ coherent state technology as reviewed in appendix \ref{CS}. Then, as was previously shown in \cite{holo2, spinor}, we obtain:
\be \label{star delta}
\delta(g)=
\f1{\pi^2}
\int d^4z\,(\la z|z\ra-1)e^{-\la z|z\ra}\,\,
e^{\la z|g|z\ra} = \f1{\pi^2}
\int d^4z\,(\la z|z\ra-1)e^{-\la z|z\ra}\,\,
E_g(z).
\ee

\medskip

The Haar measure $dg$ on $\cC(\SU(2))$ allows to determine the standard scalar product $\la,\ra_{\SU(2)}$.  The Fourier transform  should define by construction an isometry between $(\cC(\SU(2)), \la,\ra_{\SU(2)})$  and $\cC_\star(\R^3)$ equipped with a scalar product $\la \hat \phi,\hat \psi \ra= \int d\mu(z)\, \left[
(\overline{\hphi}\star\hpsi)(z)\right]$ built from a measure $d\mu(z)$, which we determine now.
\bes
\int dg\,\overline{\phi}(g)\psi(g) &\,=\,& \int dg_1dg_2\,\overline{\phi}(g_1)\psi(g_2)\delta(g_1\mone g_2) \,=\,
\f1{\pi^2} \int dg_1dg_2\,\overline{\phi}(g_1)\psi(g_2)
\int d^4z\,(\la z|z\ra-1) e^{-\la z|z\ra}\,\, E_{g_1\mone g_2}(z)\nn\\
&\,=\,& \f1{\pi^2}
\int d\mu(z)\,\left[(\overline{\hphi}\star\hpsi)(z)\right].
\ees
The measure is therefore\footnotemark:
\beq\label{measure}
d\mu(z)\,\equiv\,\frac{1}{\pi^2}d^4z (\la z|z\ra -1) e^{-\la z|z\ra}\, \Leftrightarrow [dX]= \frac{1}{\pi^2} d^3X \frac{\ka |\vec X| -1}{\ka |\vec X|} e^{-\ka |\vec X|}.
\eeq
\footnotetext{
The curious feature of this scalar product is that the measure factor $(\la z|z\ra-1)e^{-\la z|z\ra} $ on the space of spinors is not positive. However, this deviation from the Gaussian measure does not mean that the norm of functions of $z$ will be possibly negative. Indeed, since the scalar product defined with this measure factor and the $\star$-product is strictly equal to the standard scalar product between functions on $\SU(2)$, the norm will always be strictly positive unless the function vanishes.}
With all this in hand, we infer  that the Fourier transform is well-defined  in the sense that it takes  a function in $L^2(\SU(2), dg)$ to a function in $L_\star^2(\R^3,d\mu(z)) $.

\medskip

Since the Fourier transform is an isometry, we can define the inverse  Fourier transform:
\bes \label{inverse fourier}
\cF\mone: \cC_\star(\R^3)&\dr& \cC(\SU(2))\nn\\
\hf(z)&\mapsto &f(g)=
\int d\mu(z)\,
\left[e^{\la z|g^{-1}|z\ra}\,\star\hf(z)\right]=\f{\ka^3}{4\pi}\int[dX]\, e^{-\ka |\vX|} \left[
\,(E_{g\mone}\star\hat f)(\vX)\right]\,.
\ees

\medskip

A last remark on the definition of this $\SU(2)$ group Fourier transform is on taking its complex conjugate:
\be
\overline{\hf(z)}
=
\int dg\, \overline{f(g)}\,e^{\la z|g^{-1}|z\ra}
=
\int dg\, \overline{f(g^{-1})}\,e^{\la z|g|z\ra},
\ee
so that a real Fourier transform $\hf(z)\in\R$ is equivalent to $\bar{f}(g)=f(g^{-1})$.

\subsection{Defining the $\delta$-distribution on the non-commutative space}


An important missing ingredient is the non-commutative delta function $\delta^\star(\vec X)$  over $\R^3$. As usual we can determine it as a superposition of the plane-wave.
\beq\label{delta 0}
\deltas(\vec X)= \int dg\, E_{g}(\vec X)= e^{-\la z|z\ra} \one_\star= e^{-{\ka}{|\vec X|}} \one_\star.
\eeq
This shows that with this choice of parametrization the delta-distribution is actually regularized as a Gaussian  in the spinor variables. This definition extends to the case where the delta-function projects over an arbitrary point $\vec X_2$:
\bes
\deltas_{\vec X_2}(\vec X_1)
&\equiv&
\int dg \, E_{g}(\vec X_1) E_{g\mone}( \vec X_2)=  \int dg\, e^{\la z_1|g|z_1\ra}e^{\la z_2|g\mone |z_2\ra}\nn\\
&=& \f{I_1(2|\la z_2|z_1\ra|)}{|\la z_2|z_1\ra|}= \f{I_1(2|\la z_2|z_1\ra|)}{|\la z_2|z_1\ra|}e^{-\la z_1|z_1\ra} \one_\star,
\ees
where $z_1$ is the spinor for $\vec{X}_1$ and $z_2$ the spinor for $\vec{X}_2$.  Also we have made apparent the identity $\one_\star$ in the $\vec X$ variable in the last equality. We can actually relate the scalar product $\la z_2|z_1\ra$ between the spinors to the one between the vectors:
\be
|\la z_2|z_1\ra|^2
\,=\,
\tr\,|z_2\ra\la z_2|\,|z_1\ra\la z_1|
\,=\,
\f{\ka^2}2\,(|\vec X_1||\vec X_2|+\vec{X}_1\cdot\vec X_2 )\,,
\ee
which simply vanishes if $\vec X_2=0$.
In particular, when $\vec  X_2=\vec 0$, i.e. $z_2=0$, the previous definition of $\deltas_{\vec  X_2}$ gives  back \eqref{delta 0} as expected. However, it is important to notice that $\deltas_{\vec  X_2}(\vec X_1)$  is different from the more na\"ive definition $\deltas(\vec X_1-\vec  X_2)=\int dg\, E_g(\vec X_1-\vec  X_2)$, since the plane-wave is not linear in $X_1$ and $ X_2$.
Nevertheless, this delta-function $\deltas_{\vec  X_2}(\vec X_1)$ satisfies the usual properties of the delta function since $\forall \hat f\in \cC_\star(\R^3)$,
\bes
\int [dX]\, \deltas(\vec X)&=& \frac{1}{\pi^2}\int d^4z dg\, (\la z| z\ra -1)e^{-\la z| z\ra} e^{\la z|g|z\ra}= \int dg\, \delta(g)= 1,\label{norm delta}\\
\int [dX_1]\, (\deltas_{\vec X_2} \star \hat f)(\vec X_1) &=& \int [dX_1]\,  (\hat f \star \deltas_{\vec X_2})(\vec X_1)=\hat f(\vec X_2)  \label{prop delta}
\ees

It is furthermore very interesting that the expression of the delta function  $\deltas_{\vec X_2}(\vec X_1)$   defined in terms of the $z_i$ variables can be related to the notion of coherent intertwiners as introduced in \cite{un3}. Indeed, as we recall in the appendix \ref{CS}, a $n$-valent coherent intertwiner $|\{z_i\}\ra$  is given by
\beq
|\{z_i\}\ra
\,\equiv\,
\sum_{\{j_i\}}\f{1}{\prod_i \sqrt{(2j_i)!}}\,\int dg\, \bigotimes_i g\,|j,z_i\ra\,, \quad i=1,..,n,
\eeq
where the $|j,z_i\ra$ are the $\SU(2)$ coherent states following the conventions of \cite{un3,holo2, holo1}.
From this definition, we see that the norm of this coherent intertwiner gives the integral over $\SU(2)$ of products of $n$ plane-waves $E_g(z_i)$:
\be
\la \{z_i\}|\{z_i\}\ra
\,=\, \int dg\,\prod_i^n E_g(z_i).
\ee
This norm was fortunately already computed explicitly in \cite{holo2, holo1} :
\bes\label{norm}
\la \{z_i\}|\{z_i\}\ra
&\,=\,&
\sum_{J\in\N}\f{(\det \Omega)^J}{J!(J+1)!}
\,=\,
\f{I_1\left(2\sqrt{\det\Omega}\right)}{\sqrt{\det\Omega}},\\
\textrm{with}\quad
\Omega=\sum_i |z_i\ra\la z_i|,&\quad&
\det\Omega= \f{\ka^2}{2^2}\,\left[\left(\sum_i |\vec X_i|\right)^2-\left|\sum_i \vX_i\right|^2\right]\ge 0\,, \quad i=1,..,n.\nn
\ees
We notice that the norm is maximal when $\sum_i \vX_i=\vec 0$, ie the closure constraint is satisfied or equivalently $\sum_{i=1,..,n}|z_i\rangle\la z_i|\propto \id_2$. Moreover, the norm becomes more and more peaked around this maximal value in the classical limit as $\ka$ grows to $\infty$. In that sense, the integral $\int dg\,\prod_i^n E_g(z_i)$ can be interpreted as defining a smooth delta function ${\delta}_\ka$  peaked around the closure $\sum_i \vX_i=0$~:
\beq\label{delta general}
{\delta}_\ka (\vX_1,..,\vX_n)\equiv  \int dg\,\prod_i^n E_g(z_i)= \la \{z_i\}|\{z_i\}\ra.
\eeq
In particular, in the case of the bivalent intertwiner, when $n=2$, this reduces to our previous definition of $\delta^\star_{\vX}$. Indeed, we have:
$$
\deltas_{\vec X_2}(\vec X_1)
\,=\,
\int dg \, E_{g}(\vec X_1) E_{g\mone}( \vec X_2)
\,=\,
\int dg \, E_{g}(\vec X_1) E_{g}( -\vec X_2)
\,=\,
{\delta}_\ka (\vX_1,-\vX_2).
$$
We find  very interesting that the non-commutative delta function we constructed can be defined in terms of loop quantum gravity tools. This is an another example of the interplay between structures of non-commutative geometry and of loop quantum gravity  \cite{starspin}.

Notice nevertheless that the delta function ${\delta}_\ka(\vX_1,..,\vX_n)$ is not in general a straightforward function of $\sum_i\vX_i$ due to the non-linearity of the plane-wave. We can see from the explicit expression that it also depends on the total norm $\sum_i|\vX_i|$, which can not be simply factored out of the formula.
%

\medskip
Using this delta function $\deltas_{\vec X_2}(\vec X_1)$ as well as the delta function on the group $\delta(g)$, it is then straightforward to check explicitly that $\cF \circ \cF\mone= \one_{\cC_\star (\R^3)}$ and $ \cF \mone \circ \cF =\one_{\cC(\SU(2))}$, where $\cF$ is the Fourier transform.

\subsection{On the choice of plane-wave: using normalized plane-waves}
\label{normalized}\label{kernel}

\begin{figure}
\begin{center}
\includegraphics[scale=.4]{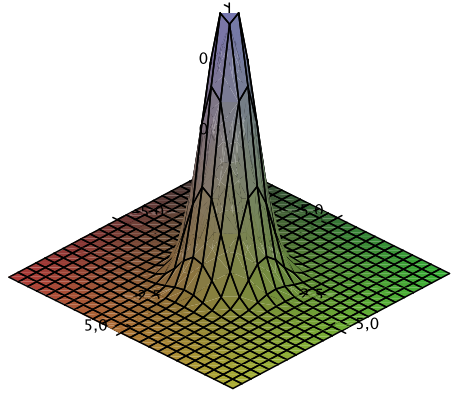}  \label{delta pic}
\caption{The (2d realization of the)  delta function $\tdelta_\ka(\vec X_1,-\vec X_2)$ peaked on  $\vec X_2=(1,1)$.}
\end{center}
\end{figure}

The $\star$-product representation of a non-commutative algebra is a highly non-unique representation. There exists actually many different star products which can be introduced through different choices  of momentum variables or more generally different choices of plane-waves.
%
We can thus change our plane-waves $E_g(z)=e^{\la z|g|z\ra }$ for other ($\U(1)$-invariant) functions of the spinor $z$. The general construction is described in appendix \ref{kernel}. Here we would like to focus on a particular choice of normalized plane-waves so that the identity for the star-product remains the trivial constant function on $\R^3$.
To this purpose, we rescale the plane-waves $E_g(z)$ by an appropriate factor:
\beq\label{normalized planewave}
\tilde E_g(z)= e^{-\la z |z\ra } e^{\la z|g|z\ra } = e^{\la z|g-1|z\ra }  \Leftrightarrow \tE_g(\vec X)= e^{\f\ka2 |\vec X|(-1+\tr g)+ \f\ka2 \tr g X}.
\eeq
We note in fact that the normalizing Gaussian factor is already present in the integral so that the delta-function over the group is
\beq
\delta(g)=\f1{\pi^2}
\int d^4z\,(\la z|z\ra -1) e^{-\la z|z\ra}\,
e^{\la z|g|z\ra}= \f1{\pi^2}
\int d^4z\,(\la z|z\ra-1) \tilde E_{g}(z).
\eeq
In this normalized case, the $\star$-product becomes
\bes \label{normalized star}
(\tilde E_{g_1}\star \tilde E_{g_2}) (z)\equiv \tilde E_{g_1g_2}(z)\Leftrightarrow ( e^{\la z|g_1-1|z\ra })\star   e^{\la z|g_2-1|z\ra })\equiv    e^{\la z|g_1g_2-1|z\ra }
\Leftrightarrow (\tilde E_{g_1}\star \tilde E_{g_2}) (\vec X)\equiv \tilde E_{g_1g_2}(\vec X)
\ees
The previous construction of the Fourier transform goes along the same way as in the previous section, but the identity is now trivial, $\one_\star=\one$. The delta-function over configuration space is given by the Gaussian in the spinor variables, or in the $\vec X$ variables as
\beq
 \tdelta (\vec X)= e^{-\ka |\vec X|}\,.
\eeq
The generalized delta functions $\widetilde{\delta}^\star(\vX)$, and more generally the distributions $\widetilde{\delta}_\ka(\vX_1,..,\vX_n)$, are now given as a function of $z_i$ as
\beq\label{delta general normalized}
\widetilde{\delta}_\ka(\vX_1,..,\vX_n)=  \frac{\la \{z_i\}|\{z_i\}\ra}{\prod_{i}{e^{\la z_i|z_i\ra}}}.
\eeq
Their  expressions in terms of  $X_i$ can be easily read from \eqref{norm}. Figure (\ref{delta pic}) illustrates the shape of $\widetilde{\delta}_\ka(\vX_1,-\vX_2)$.

\subsection{Computing the $\star$-product}
\label{proprietes}


It is now natural to ask what is the structure of the $\star$-product we have introduced.
Although we study here the star-products induced by the choices of plane-waves $E_g$ and $\tE_g$, the discussion below applies to all plane-waves of the type $\cK^*_g$ and their induced $\star$-products as introduced in appendix \ref{kernel1}.
We first analyze the  $\star$-product between coordinates to check that we recover the $\su(2)$ non-commutative structure.  

\medskip

Calling now $p^i$ and $q^i$ the coordinates for respectively the group elements $g_1$ and $g_2$ such that
\beq
 g_1g_2 \rightsquigarrow \vp\oplus\vec q= \sqrt{1-\f{\vec q^2}{\kappa^2}}  \, \vec p +   \sqrt{1-\f{\vec p^2}{\kappa^2}}  \, \vec q - \frac{1}{\ka} \vec p\wedge \vec q,
\eeq
the $\star$-product  between coordinates for the  plane-waves $E_g$ is then \footnotemark:
\bes
X_i \one_\star \star X_j\one_\star&=& -\int dg_1dg_2\, \delta(g_1)\delta(g_2)\partial_{p^i} \partial_{q^j} E_{g_1g_2}(z)
=  \left(X_iX_j + \frac{1}{\ka}\left( \delta_{ij}|\vec X|+  i\epsilon_{ij}^k X_k\right) \right) \one_\star\,,
\label{st prod X}
\ees
\footnotetext{
To prove this we have used the following identities
\bes
\partial_{q^j}\la z|g_1g_2|z\ra_{|_{\vec q=\vec p=\vec 0}}&=& \partial_{p^j}\la z|g_1g_2|z\ra_{|_{\vec q=\vec p=\vec 0}}= i  X_j \nn\\
\partial^2_{p^iq^j}\la z|g_1g_2|z\ra_{|_{\vec q=\vec p=\vec 0}}&=&-\kamone \left(\delta_{ij}|\vec X| + i\epsilon_{ij}^k X_k\right)\nn
\ees}
where $\one_\star=e^{\ka |\vec X|}$. It is interesting to notice that the previous formula \eqref{st prod X} is actually very similar to the one derived by the authors of \cite{fourier2} and obtained when considering the 4d Fourier transform for $\SU(2)$, with $T=|\vec X|$ in their notations.

In the case of the plane-waves $ \tilde  E_g(z)$, the non-trivial identity $\one_\star$ drops out and we have more simply:
\bes
X_i  \star X_j &=&X_i X_j+\frac{1}{\ka}\left( \delta_{ij}|\vec X|+  i\epsilon_{ij}^k X_k\right). \label{nice}
\ees

In both cases, it is direct to see
that the $\star$-product we introduced is in fact a realization of the $\su(2)$ non-commutative structure since
\beq
[X_i\one_\star ,X_j\one_\star ]_\star= X_i\one_\star \star X_j\one_\star- X_j\one_\star\star X_i\one_\star= \frac{2}{\ka} i \epsilon_{ij}^k X_k\one_\star\,.
\eeq

\medskip

Earlier we have defined $\cC_\star(\R^3)$ to be the functions which are generated by the $X_i(z)$ and as such invariant  under the phase rescaling $z\dr e^{i\alpha} z$. This algebra can also be characterized as the algebra generated by the functions which $\star$-commute with $|\vec X|$.  Indeed, we have
\beq
|\vec X| \one_\star \star X_i \one_\star=   (|\vec X| X_i + \frac{1}{\ka} X_i) \one_\star= X_i\one_\star \star |\vec X|\one_\star,\qquad
[|\vec X|\one_\star, X_i\one_\star]_\star =0\,.
\eeq
From this we deduce that any function built out from the $X_i$ will $\star$-commute with $|\vec X|$.

We shall show in section \ref{voros} that these two characterizations of $\cC_\star(\R^3)$ are indeed equivalent.

\medskip




\section{$\su(2)$ non-commutativity from the quantum oscillator perspective}

In the previous section we have shown that the spinorial approach allows to introduce a well-defined Fourier transform for $\SU(2)$. In this section, we want to show that this approach allows to shed a new light on the $\su(2)$ non-commutative structure. First, we show that the 4d (bicovariant) differential calculus on $\su(2)$ can be naturally recovered from this approach. Second, we prove that the Voros star-product between the spinor variables gives exactly the $\SU(2)$ $\star$-product of the previous section based on the normalized plane-waves $\tE_g(z)$.
This provides a simple representation of our $\star$-product as a differential operator. 

\subsection{$\su(2)$ bicovariant differential calculus}

As we have recalled in section \ref{spinor}, the spinorial approach developed in \cite{un1, un2, un3, un4} relies on the Schwinger representation of $\su(2)$. More explicitly the spinor variables $z\in\C^2$ are quantized, ie $z_a\dr a_a$, $\overline{z}_a\dr a_a^\dagger$
\beq
\{z_a,\overline{z}_b\}= -i \delta_{ab} \dr [a_a, a_b^\dagger]= \delta_{ab},
\eeq
and the dimensionful $\su(2)$ generators are simply
\beq
\vec \XX= \kamone  a ^\dagger _a \vsigma _{ab} a_b,
\qquad \ka |\vec \XX|= \sum_a a^\dagger _a a_a=\sum_a N_a,
\eeq
where $N_a$ is the number operator.
Using the commutation relations of the creation/annihilation operators, we recover
\beq
[\XX_i,\XX_j]=\f2\kappa i \epsilon_{ij}^k\XX_k, \quad [\sum_a N_a,X_j]=0.
\eeq
The non-commutative space $\hat\R^4\sim \hat\R^2\times \hat\R^2$ generated by the operators $ \alpha_\mu=(a_0, \, a_0^\dagger,\, a_1, \, a_1^\dagger)$ is equipped with a differential structure  which satisfies the Leibniz law,
\bes
d(\alpha_\mu \alpha_\nu)= (d\alpha_\mu)\alpha_\nu+ \alpha_\mu(d\alpha_\nu), \quad  \forall \mu,\nu,
\eea
and such that the 1-forms commute with $\alpha_\mu$
\beq
[d\alpha_\mu, \alpha_\nu]=0.
\eeq
This last property can be also understood as the fact that the translation symmetry is not deformed in this case\footnote{It is only when a Poincar\'e transformation is performed -- that is in the relevant 2d Euclidian case, both a rotation and a translation -- that the differential structure is non-trivial. In fact there is no pure rotation transformation. For further details see \cite{giovanni}.}.

Using this, we can now calculate in a direct manner, the commutators of $\XX_i$ and $d\XX^j= da^\dagger_b\sigma^j_{bc} a_c+ a^\dagger_b\sigma^j_{bc} da_c$.
\bes
[\XX_i, d\XX_j] &=&\frac{1}{\ka^2}\left(-i\epsilon_{ilj}\left(da^\dagger_a\sigma^l_{ab} a_b+ a^\dagger_a\sigma^l_{ab} da_b\right)+ \delta_{i}^j\sum_c(a^\dagger _c da_c- a_c d a^\dagger_c)\right) \nn\\
&=& \kamone \left( i\epsilon_{ij}^k d\XX_k + \delta_{ij}\Theta\right)
\ees
We see therefore that there is an extra contribution $\Theta=\kamone\sum_c(a^\dagger _c da_c- a_c d a^\dagger_c) $ that appears. This is the non-trivial fourth component of the $\su(2)$ bicovariant differential calculus \cite{batista}. Using again the quantum  harmonic  oscillator expressions, we get
\beq
[\Theta, \XX_i]= -\frac{1}{\ka}d\XX_i,
\eeq
which is again consistent with \cite{batista}.

\subsection{Voros $*$-product versus $\SU(2)$ $\star$-product}\label{voros}

The quantum harmonic oscillators can naturally be defined in the Moyal representation of quantum mechanics \cite{moyal}. More precisely, we start  with $c$-numbers, here the spinorial variables $z$, and we introduce a $*$-product encoding the quantum structure and realizing an exact quantization through the Weyl map:
\beq\label{moyal}
 [a_i, a_i^\dagger]= \delta_{ij}\dr [z_i,\overline{z}_j]_*= z_i*\overline{z}_j-\overline{z}_j*z_i=  \delta_{ij}.
\eeq
There exist many realizations of such \stpm, the most well-known are
the Moyal product  and the Voros product defined respectively as:
\bes
(f_1*_mf_2)(z)&=&f_1(z) e^{\demi(\overleftarrow{\pp_z}\overrightarrow{\pp_{\overline{z}}}-\overleftarrow{\pp_{\overline{z}}}\overrightarrow{\pp_z})} f_2(z),\label{moyal prod}\\
(f_1*_v f_2)(z)&=&  f_1(z)e^{\overleftarrow{\pp_z}\overrightarrow{\pp_{\overline{z}}}}f_2(z).\label{voros prod}
\ees
Both of these $*$-products allow to recover the commutation relation in \eqref{moyal}.
%
Let us point out that these two *-products define unitary-equivalent quantization maps, the Moyal product corresponding to the Weyl symmetric ordering while the Voros product corresponds to the normal ordering.


Following this logic,
we would like to re-express the Schwinger representation as $\XX_i \dr X_i= \frac{1}{\ka} \overline{z}_a \vec \sigma z_b$ using a Weyl map and define the relevant $*$-product between the spinors  $z$ to calculate the commutators of the $X_i$, seen as a function of the spinor variables $z$, such that we get:
\bes
[X_i,X_j]_* &=&\f2\kappa \epsilon_{ij}^kX_k, \quad   [|\vec X|, X_i]_*=0.
\eeq
A natural question is now to wonder if one such $*$-product between the spinorial variables is actually equivalent to the $\SU(2)$ $\star$-product that we have defined in the previous section \ref{proprietes}.
The Moyal and Voros products give  respectively
\bes
X_i*_m X_j= X_iX_j + \frac{1}{\ka}\epsilon_{ij}^k X_k,\\
X_i*_v X_j= X_i X_j+\frac{1}{\ka}\left( \delta_{ij}|\vec X|+  i\epsilon_{ij}^k X_k\right).
\ees
We see therefore that the Voros $*$-product gives the same star-product between coordinates as our one normalized $\star$-product considered in \eqref{nice}:
\beq
X_i\star X_j = X_i*_vX_j.
\eeq
Let us extend this analysis to the plane-waves $\tE_g(z)$ and calculate the Voros $*$-product between them in order to fully check that we recover the $\star$-product. Writing the normalized plane-waves as $\tE_g(z)= e^{-\la z|z\ra}e^{\la z | g|z\ra}=e^{\la z | g-1|z\ra}$ in terms of the spinors $z$ proves efficient and we compute:
\bes
(\tilde E_{g_1} *_v \tilde E_{g_2}) (z)&=& \left( e^{\la z | g_{1}-1|z\ra}\right) e^{\overleftarrow{\pp_z}\overrightarrow{\pp_{\overline{z}}}} \left( e^{\la z | g_{2}-1|z\ra}\right)\nn\\
&=& \left( e^{\la z | g_{1}-1|z\ra } e^{(\la z | (g_{1}-1))\cdot  \overrightarrow{\pp_{\overline{z}}}}\right) \left( e^{\la z | g_{2}-1|z\ra}\right)=   e^{\la z | g_{1}-1|z\ra }   e^{\la z | g_{2}-1|z\ra }    e^{\la z | (g_{1}-1)(g_2-1)|z\ra } \nn\\
&=& \left( e^{\la z | g_{1}g_2-1|z\ra}\right) = \tilde E_{g_1g_2}(z)= (\tilde E_{g_1} \star \tilde E_{g_2}) (z)
\ees
This shows explicitly that the Voros $*$-product reproduces exactly our $\SU(2)$ $\star$-product defined in \eqref{normalized star} and provides a proper representation of the $\su(2)$ non-commutative structure:
\beq
(\hphi \star \hpsi)(z)=(\hphi *_v \hpsi)(z), \quad \forall \hpsi,\, \hphi\in \cC_\star{(\R^3)}.
\eeq
Furthermore, using the definition for the Voros $*$-product and
$$
\partial_{z_a}= \kamone \overline{z}_m\sigma^i_{ma}\frac{\pp}{\pp X_i}, \quad \partial_{\overline{z}_a}= \kamone \sigma^i_{an} z_n \frac{\pp}{\pp X_i},
$$
we can use this equality to give a nice and simple expression for the $\SU(2)$ $\star$-product \cite{moyal1},
\beq
(\hphi \star \hpsi)(\vec X)
\,= \,
\hphi(\vec Y)
e^{\kamone\left(|\vec X|\delta_{ij} +i \epsilon_{ij}^k X_k\right) \overleftarrow{\pp_{Y_i}}\overrightarrow{\pp_{Z_j}} }
\hpsi(\vec Z)_{|_{Y=Z=X}}, \quad \forall \hphi, \hpsi\in\cC_\star(\R^3)\,.
\eeq
This provides us with an expression as a differential operator for the $\SU(2)$ $\star$-product based on the normalized plane-waves $\tE_g$. By re-inserting the Gaussian normalization, we can easily deduce from it a differential operator representation for the $\SU(2)$ $\star$-product based on the original spinorial plane-waves $E_g$, which will be nevertheless less elegant.

\medskip
Thanks to the Voros realization of the $\SU(2)$ $\star$-product, we can re-examine the equivalent definitions of the $\cC_\star (\R^3)$. We have seen it is given by the subalgebra of functions of $\cC(\C^2)$ which $\star$-commute with $|\vec X|$. Using the Voros representation, this becomes then
\beq\label{voros norm}
{[} |\vec X|, f(z,\overline z) {]}_{\star_v} = 0 \Leftrightarrow \left( \overline{z}\partial_{\overline{z}} - z\partial_z\right)f(z,\overline z) =0.
\eeq
On the other hand, $\cC_\star(\R^3)$ is generated by the functions which are invariant under the rescaling $z\dr e^{i\alpha}z$, that is
\beq\label{equal}
f(z,\overline{z})= f(e^{i\alpha}z,e^{-i\alpha}\overline{z}), \quad \forall f\in \cC_\star (\R^3).
\eeq
Consider a small $\alpha$ and expand \eqref{equal}, we get
\beq
- i\alpha\left( \overline{z}\partial_{\overline{z}} - z\partial_z\right)f(z,\overline{z}) =0,
 \eeq
which is equivalent to \eqref{voros norm}.

\medskip

We can recap the present situation: there are different ways to construct a star product representation of a non-commutative structure. One consists in identifying the momentum addition structure and from this infer the star product between coordinates using a Fourier transform. This is the approach we followed in section \ref{su2}. Another one consists in defining a Weyl map by brute force. In this section we have constructed such Weyl map, starting from the Voros representation of the quantum oscillators. We have shown that the two representations are the same.

\smallskip

As a final comment, we note that considering the momentum structure, and in particular the delta function over momentum space allows to recover the right measure in the configuration space. We have seen that in the case of $\tilde E_g(z)$, the relevant measure is given by $d\mu(z)= d^4z (\la z| z \ra -1)$. In \cite{moyal1}, the authors did not consider the momentum structure and therefore only took the standard measure $d^4z$, which is not the correct measure, as we have shown.



\section{NCQFT Representation of Group Field Theory}

Spinfoam models (for quantum gravity and topological field theories) can be defined at the non-perturbative level through group field theories (GFTs), which are non-local field theories defined on Lie group manifolds (for a review, see e.g. \cite{review-gft}). For instance, the most studied case is the Boulatov group field theory for 3d Euclidean quantum gravity \cite{boulatov}. It is indeed the standard model to discuss   the issue of renormalization in the context of GFT before addressing the more complicated GFT's describing spinfoam models for 4d gravity. The model is formulated on the manifold $\SU(2)^3$, satisfies a $SU(2)$-gauge invariance and has a non-local interaction term.

In the previous section, we have defined a $\SU(2)$ $\star$-product and shown its equivalence to the Voros *-product in the spinor variables. This opens new possibilities. Indeed, we can use our new $\SU(2)$ Fourier transform and write the GFT in terms of the spinor variables. As we have seen, the  non-commutativity is then a ``standard" one, of the Moyal type, and we hope to be able to use standard renormalization techniques already developed for non-commutative quantum field theories based on the Moyal and Voros $*$-products.

In this section we will describe explicitly Boulatov's model in terms of the spinor variables and discuss the realization of the quantum symmetries given by the quantum double DSU$(2)$. As a warm up, we first consider the 2d GFT on $\SU(2)$ \cite{gftmatter1,gftnc1}.
%

\subsection{2d GFT on $\SU(2)$}

We consider a (real) scalar field theory on $\SU(2)$ with a field $\phi\in\cC(\SU(2))$ and action is given by:
\be
S_{2d}[\phi]=\f12\int dg\, \phi(g_1)\phi(g_2)\delta(g_1g_2)
+\sum_{n\ge 3}\f{\alpha_n}{n!}\int [dg]^n\, \phi(g_1)..\phi(g_n)\,\delta(g_1..g_n)\,.
\ee
From the perspective of non-commutative field theories,  $\SU(2)$ is the momentum space and  the term $\delta(g_1..g_n)$ corresponds to the  conservation law of momenta. We can now use  the normalized plane-wave $\tilde E_g(z)$, to implement our Fourier transform and express the action $S_{2d}[\phi]$ in configuration space. We consider therefore $\hphi\in\cC_\star(\R^3)$ with
$$
\hphi(z)=\int dg\,\phi(g)\,\tE_g(z)\qquad
\tE_g(z)\,=\f{e^{\la z|g|z\ra}}{e^{\la z|z\ra}},\qquad
\tE_{g_1}\star\tE_{g_2}=\tE_{g_1g_2}\,.
$$
A straightforward implementation of the Fourier transform gives
\be
S[\phi]=\f12\int d\mu(z)\,(\hphi\star\hphi)(z)
+\sum_{n\ge 3}\f{\alpha_n}{n!}\int d\mu(z)\,\hphi^{\star n}(z)\,,
\ee
with the measure $d\mu(z)=(\la z|z\ra -1)d^4z$. As we have shown the equivalence of the   $\star$-product and the Voros $*$-product, this action can also be written as
\be\label{2d action 1}
S[\phi]=\f12\int d\mu(z)\,(\hphi*_v\hphi)(z)
+\sum_{n\ge 3}\f{\alpha_n}{n!}\int d\mu(z)\,\hphi^{*_v n}(z)\,.
\ee
We  recall that unlike the Moyal star product, we have
\beq
\int dz\, (\hphi*_v\hpsi)(z)\neq \int dz\, \hphi(z)\hpsi(z).
\eeq
The field theory defined by $S_{2d}[\phi]$ can also be written as a (sum over) matrix model \cite{gftnc1}. We have shown here that we can write this GFT as a Voros non-commutative field theory, with a scalar field $\hphi(z)\in\cC(\R^3)$ invariant under $z\dr e^{i\alpha}z$. It would be interesting to explore further the properties of this new formulation, especially with respect of the renormalization of the theory.  In any case, the immediate advantage of our formalism over the previous works \cite{gftnc1,gftnc2,gftsym1,gftsym2} is that we are truly dealing with a field living in $\SU(2)$ and we do not restrict ourselves to even fields living only in $\SO(3)$. At the spinfoam level, this means that we will get all $\SU(2)$-representations of arbitrary spin $j\in\N/2$ and not only even representations with integer spins.

\medskip

As soon as we have a law of conservation of momenta, one expects some translational symmetry involved. Indeed there exists an action on our field $\phi\in\cC(\SU(2))$ of the quantum double $D\SU(2)$, which is a deformation of the Euclidian group $ISO(3)$ \cite{fourier3}. We emphasize again that our spinor formalism allows   for the full action of $D\SU(2)$ and not only $D\SO(3)$. The quantum double is given as an algebra in terms of the cross-product between the algebra of functions of $\cC(\SU(2))$ and the group algebra $\C\SU(2)$, as   $D\SU(2)= \cC(\SU(2))\rtimes \C\SU(2)$ \cite{majid}. The action of $D\SU(2)$  on $\phi\in \cC(\SU(2))$ is given by the action of translations parameterized by   an arbitrary spinor $ a\in\C^2$ and the action of rotations parameterized by a group element $u\in\SU(2)$  which are respectively
\bes\label{sym dsu2}
\phi(g)&\dr& \tE_g(a)\phi(g), \quad
\phi(g_1)\otimes \phi(g_2)\dr
(\tE_{g_1}(a)\star\tE_{g_2}(a))\,\phi(g_1)\otimes \phi(g_2)=\tE_{g_1g_2}(a)\phi(g_1)\otimes \phi(g_2),
\quad a\in \C^2, \label{translation}\\
\phi(g)&\dr&  \phi(ugu\mone),\quad  \phi(g_1)\otimes \phi(g_2)\dr  \phi(ug_1u\mone)\otimes \phi(ug_2u\mone) \quad u\in \SU(2). \label{rotation}
\eeq
It is not difficult to check that due to the conservation of momenta, the action $S_{2d}(\phi)$ is invariant  under such translations, and that thanks to the Haar measure and the invariance of  $\delta$ under rotations, $S_{2d}(\phi)$ is also invariant under the rotations.

\medskip

A natural question to explore is the realization of the $D\SU(2)$ symmetries in configuration space, i.e. the analogue of \eqref{translation} and \eqref{rotation} for $\cC_\star(\R^3)$. To this purpose, we  perform the Fourier transform of  \eqref{translation} and \eqref{rotation} . The rotations simply read:
\beq
\phi(ugu\mone)\,\mapsto\,  \hphi(uXu\mone),
\eeq
and we recover the adjoint action of $\SU(2)$ on the coordinates $\vec X$, as expected. On the other hand, the translations are trickier. In that case, the Fourier transform  reads:
\bes\label{fuzzy translation}
\tE_g(a)\phi(g)\,\mapsto\,
\int dg\, \phi(g) \tE_{g}(z_1) \tE_g(a)&=&\int dg d\mu(z_2)\, \hphi(z_2)\star \tE_{g\mone}(z_2)\tE_{g}(z_1) \tE_g(a) \nn\\
&=&\int  d\mu(z_2)\, \hphi(z_2)\star \widetilde\delta_\ka(\vec X_1,- \vec X_2,\vec A).
\ees
We obtain a ``fuzzy" implementation of translations in configuration space, with the generalized delta function $\delta_\ka$ defined in \eqref{delta general normalized}. It is not an exact translation $\vec X\dr \vec X+ \vec A$, essentially due to the non-linearity of the plane-wave.

%

\subsection{Boulatov model}
\bigskip

We can easily adapt the procedure described above  to the 3d case. Indeed, let us consider the colored Boulatov GFT \cite{colorGFT,gftsym1,gftsym2,gftnc2} with complex fields $\phi_{c=1..4}\in \cC(\SU(2)^3)$ which are right translational invariant
\beq
\phi_c(g_1,g_2,g_3)\equiv\int dh\, \phi_a(g_1h,g_2h,g_3h), \quad  \forall c=1..4.
\eeq
The colored Boulatov action is given by:
\bes
S_b[\phi_c]
&=&
\f12\int [dg]^3\,\sum_c^4\phi_c(g_1,g_2,g_3) \overline \phi_c(g_1,g_2,g_3)\\
&&+\f\alpha{4!}
\int [dg]^6\,
\phi_1(g_1,g_2,g_3)\phi_2(g_3,g_4,g_5)\phi_3(g_5,g_2,g_6)\phi_4(g_6,g_4,g_1)\, + c.c.. \nn
\ees
This GFT generates the Ponzano-Regge amplitudes for Euclidian gravity, as it is easy to see by looking at the Feynman amplitudes and using the Peter-Weyl theorem. This same field theory defined over $\SO(3)^3$ has been recently studied  using non-commutative techniques based on the plane-waves $e^{\tr |g|X}$ discussed in section \ref{so3}. This approach allowed on one hand to connect GFT with simplicial geometry, since the non-commutative variable $X$ can be interpreted as discretized $B$-field \cite{gftnc2} (also see \cite{Bobs} for a discussion on the extent of the validity of the identification of $X$ as the discretization of the $B$ field) and on the other hand to connect the quantum group symmetries of the GFT to the diffeomorphism symmetry of the $BF$ action \cite{gftsym2}.  Since the spinfoam Ponzano-Regge model is defined for $\SU(2)$, we intend now to discuss Boulatov's model on $\SU(2)$ in the light of the new spinorial Fourier transform.

\subsubsection{Non-commutative variables and discretization of the $BF$ action}
We first perform the Fourier transform on the fields $\phi_c$, and consider\footnote{The algebra $ \cC_\star({\R^3}^{\times 3})$ is seen as a subalgebra of $\cC({\C^2}^{\times 3})$. The fields $\phi_c$ are therefore  invariant under the rescaling by independent phases of the spinors: $z_i\dr e^{i\alpha_i}z_i, \, \overline{z}_i\dr e^{-i\alpha_i}\overline{z}_i$.} $\hphi_c(z_1,z_2,z_3)\in \cC_\star({\R^3}^{\times 3})$, $\forall c=1,..,4.$
\bes
\hphi_c(z_1,z_2,z_3)&\equiv& \int [dg]^3dh \, \phi_c(g_1h,g_2h,g_3h)\,\tE_{g_1}(z_1)\tE_{g_2}(z_2)\tE_{g_3}(z_3)\nn\\
&=&  \, \hphi_c(z_1,z_2,z_3)\,\star_{1,2,3} \, \int dh \tE_{h}(z_1)\tE_{h}(z_2)\tE_{h}(z_3)\\
&=& \hphi_c(z_1,z_2,z_3)\star_{1,2,3} \hat C(z_1,z_2,z_3)\,,
\ees
where $\hat C(z_1,z_2,z_3)= \frac{\la \{z_i\}|\{z_i\}\ra}{\prod_{i=1}^3{e^{\la z_i|z_i\ra}}}=\widetilde{\delta}_\ka(\vX_1,\vX_2,\vX_3)$. As we discussed earlier, $\widetilde{\delta}_\ka(\vX_1,\vX_2,\vX_3)$ is the norm of the trivalent coherent intertwinner and it defines  a smooth  delta function peaked around the closure $\sum_i \vX_i=0$.
Geometrically, as in the $\SO(3)$ case, we still interpret $\hphi_c(z_1,z_2,z_3)=\hphi_c(X_1,X_2,X_3)$ as  representing a quantized triangle where the vectors $\vX_i$ are considered as the normals to the edges and the closure of the triangle $\sum_i \vX_i=0$ is implemented in a ``fuzzy" way.

The Fourier transform can be performed  on the  Boulatov action $S_b[\phi_c]$. Since the $\star$-product is the dual of the convolution product, the combinatorial structure of the action is preserved.
Using the equivalence between the $SU(2)$ $\star$-product and Voros $*$-product, we write it as a Voros non-commutative field theory. To keep the notations simple, we define $\hphi_1(X_1,X_2,X_3)\equiv \hphi_{(123)}$, $\hphi_2(X_3,X_4,X_5)\equiv \hphi_{(345)}$ and so on and so forth. The action becomes then
\bes
S_b[\phi_c]&=&
\f12\int [dX]^3\,\sum_c^4\hphi_c{}_{(123)}* \overline \hphi_c{}_{(123)}\\
&&+\f\alpha{4!}
\int [dX]^6\,
\hphi_1{}_{(123)}*\hphi_2{}_{ (345)} * \hphi_3{}_{(526)}*\hphi_4{}_{(641)}\, + c.c., \nn
\ees
where $\hphi_{(i)}*\hphi_{(i)}\equiv( \hphi*\hpsi)(X_i)$, with $\hpsi(X_i)=\hphi(-X_i)$ and where $[dX]$ is the non-trivial measure $[dX]= d^3X\frac{|\vec X|-1}{|\vec X|}$.
It would be interesting to understand if this new formulation  can provide  new angles of attack for the renormalization analysis.

\begin{figure}
\begin{center}
\includegraphics[scale=1.8]{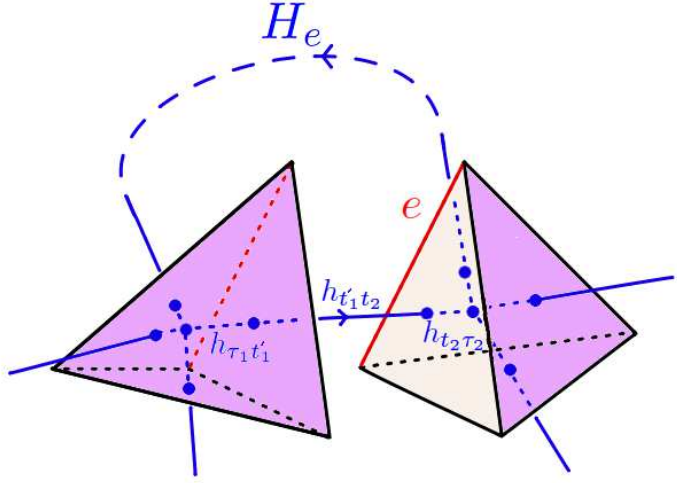}
\caption{Holonomy $H_e$ around the face dual to the edge $e$. The face $t^\prime_1$ is identified with the face $t_2$. The notation $h_{t'_1t_2}$ standing for $h_{t_2}$ allows to make obivous the orientation of the spinfoam link dual to the triangle.}\label{fig:geometry}
\end{center}
\end{figure}

\medskip
In the $\SO(3)$ case, the interpretation of the $X$ variables came when looking at the Feynman amplitudes of the GFT written in terms the configuration variables $X$. Indeed these Feynman amplitudes give the spinfoam amplitudes for BF theory which are understood as discretization of the path integral for 3d quantum gravity. This provides the Feynman amplitudes and the variables $X$ with a clear geometrical interpretation.
In the present case in the colored GFT defined on $\SU(2)$, the kinetic and interaction terms provide the following propagator $\cP(\vec X,\vec Y)$ and vertex  contribution $\cV(\vec X,\vec Y)$:
\bes
\cP(\vec{X}_1,..,\vX_3,\vec{Y}_1,..,\vec{Y}_3)
&=& \int dh_t\, \prod_{i=1}^3 \left(\tdelta_{- X_i}\star \tE_{h_{t}}\right)(Y_i), \nn \\
\cV(\vec{X}_1,..,\vX_6,\vec{Y}_1,..,\vec{Y}_6)
&=&\int \prod_t dh_{\tau t}\, \prod_{i=1}^6 \left(\delta_{- X_i}\star \tE_{h_{t \tau }h_{\tau t'} }\right)( Y_i)  \label{vertex}.  
\ees
Following \cite{gftnc2}, $t$ labels the triangles and $\tau$ the tetrahedra. The group variables $h_{t}$ and $h_{t\tau}$, with the convention $h_{\tau t}=h_{t\tau}^{-1}$,  come from the right invariance of the fields and are respectively interpreted as parallel transport respectively through the triangle $t$ and from the tetrahedron $\tau$ to the triangle $t$,  cf Figure \ref{fig:geometry}.
This is exactly the same combinatorial and algebraic structure as the one derived for the $\SO(3)$ case, expect for the fact that the plane-waves are different in the two cases. As a result, we can follow the steps of \cite{gftnc2} and we obtain the Feynman amplitude for a graph $\Gamma$:
\bes
Z(\Gamma)&=& \int \prod_t dk_t \prod_e [dX_e]\, \tilde E_{H_e}(X_e)=\int \prod_t dk_t \prod_e [dX_e]\,  e^{\frac{\ka}{2} |X_e|(-1+ \tr H_e )+\frac{\ka}{2}\tr X_e H_e}  \label{amplitude} \\
&=&  \int \prod_t dk_t \prod_e d\mu(z_e)\, e^{\la z_e|H_e-1|z_e\ra} \nn
\ees
The group element is the parallel transport associated to the triangle $t$, it is defined as the oriented product $k_t=h_{\tau t}h_t h_{t\tau'}$ where $\tau$ and $\tau'$ are the two tetrahedra sharing the triangle. Then $H_e=\overrightarrow{\prod}_{t\ni e} k_t$ is the holonomy around the dual face to the edge $e$. More precisely, labeling $t_i$ and $\tau_i$ the triangles and tetrahedra sharing the edge $e$ with $i=1..N$ (and identifying $t_{n+1}=t_1$), the holonomy reads
$H_e=\prod_{i} h_{\tau_it_{i+1}}h_{t_{i+1}}h_{t_{i+1}\tau_{i+1}}$ (see figure \ref{fig:geometry}).
This expression \eqref{amplitude} provides an expression in terms of the spinor variables $z_e$ of the path integral for the $BF$ theory over the triangulation dual to the Feynman diagram $\Gamma$. It was already derived using coherent intertwiner techniques in \cite{holo2}.
We compare it to the expression for the discretized path integral for $BF$ theory in the $\SO(3)$ case \cite{pr1,fourier0}:
\bes
Z_{\SO(3)}(\Gamma)&=& \int \prod_t dh_t \prod_e dX_e\, e_{H_e}(X_e)=\int \prod_t dh_t \prod_e dX_e\,  e^{\frac{\ka}{2}\tr X_e |H_e|}.  \label{amplitude so3}
\ees
The key differences between \eqref{amplitude so3}  and \eqref{amplitude}  lie  in the choices of measure  $dX_e$ (which is trivial for the $\SO(3)$ case) and the plane-waves.  The term ${\frac{\ka}{2}\tr X_e H_e}$  in \eqref{amplitude so3} corresponds to the ``natural" discretization  of the $BF$ action. However \textit{rigorously}, from the non-commutative point of view, this discretization looses track of the full $\SU(2)$ structure and keeps only  $\SO(3)$ as explained in section \ref{so3}.
Our spinorial approach suggests here that a ``good" discretization, which would keep track of the full $\SU(2)$, should involve a non-trivial discretization of the $BF$ action as follows:
\beq
\tr B\wedge F
\quad\longrightarrow\quad
\frac{1}{2} |X_e|(-1+ \tr H_e )+\frac{1}{2}\tr X_e H_e,
\eeq
and the associated non-trivial measure $[dX_e]= d^3X_e\frac{|\vec X_e|-1}{|\vec X_e|}$.
As we have shown, this choice allows us to recover rigourously the discretized path integral for $BF$ theory with gauge group $\SU(2)$:
\beq
Z(\Gamma)&=&\int \prod_t dh_t \prod_e [dX_e]\, \tilde E_{H_e}(X_e)
\,=\,
\int \prod_t dh_t\,\prod_e\delta(H_e), \textrm{ with } H_e\in\SU(2)\,.
\eeq

\subsubsection{Boulatov symmetries}
The invariance of the colored Boulatov action defined over $\SO(3)^{\times 3}$ under the action of four copies of $D\SO(3)$ has been explained explicitly in \cite{gftsym2}.  We are now considering the colored Boulatov model  defined over $\SU(2)^{\times 3}$ and  we can perform  a similar analysis by generalizing \eqref{translation} and \eqref{rotation} to $\DSU(2)^{\times 4}$ following the lines of \cite{gftsym2}. The symmetry analysis of Boulatov action can be performed in a momentum space given by either $\SO(3)^{\times 3}$ or $\SU(2)^{\times 3}$  in an analogous manner.  As was shown in \cite{gftsym2}, the quantum group symmetry of the interaction term  corresponds to the invariance of the spinfoam amplitudes under translating the four summits of a tetrahedron.
\smallskip

The key difference between the $\SU(2)$ and the $\SO(3)$ cases comes when implementing the symmetries at the configuration space level. Indeed the translational symmetry of Boulatov model is related to the translational symmetry of the $BF$ action thanks to the Bianchi identity \cite{pr1,pr2},
\beq
B\dr B+d_A\phi,
\eeq
where $d_A$ is the covariant derivative with respect to the connection $A$, which curvature is $F$. When discretizing the $BF$ action into $\frac{1}{2}\tr X_e\,H_e$ for $\SO(3)$, the discretized Bianchi identity still implies invariance under the transformation $X_e\dr X_e+ A_e$, which is the non-commutative realization of the translation. In the spinorial approach for $\SU(2)$, the discretized $BF$ action is $\frac{1}{2} |X_e|(-1+ \tr H_e )+\frac{1}{2}\tr X_e H_e$ and is therefore non-linear in $X$. From this perspective it is clear that the translational symmetry should be realized in a non-standard way. This is precisely what we have obtained when in \eqref{fuzzy translation}, where we have argued that the translations are implemented in a ``fuzzy" way.
\medskip

To summarize, even though the $BF$ action is discretized in a non-standard way with a non-linear term in $X_e$ (necessary to account for the full $\SU(2)$ structure), a translational symmetry still exists and is implemented in a non-trivial way.

\section*{Conclusion \& Outlook}
Let us summarize what we have done before presenting  the new directions that  our approach leads to.  Essentially we have applied the spinor representation to the GFT context. This have a number of nice implications. First, this means that we can use the non-commutative tools for a GFT defined on $\SU(2)$ and not only on $\SO(3)$. Thanks to the link between GFT and simplicial geometry, the spinor representation points towards a different discretization of the $BF$ action (as was already shown in \cite{holo2}), such that the full $\SU(2)$ structure is kept into account.
Second, we have pinpointed that the use of the spinor representation allows for  a natural derivation of the 4d structure of the bicovariant calculus on $\su(2)$.
Third, we have shown that our $\star$-product for $\SU(2)$ is given by the Voros $*$-product between the spinors, unlike the $\SO(3)$ star-product which still remains rather mysterious despite several studies.
Finally, we have discussed the implementation of the  quantum group symmetries given by $D\SU(2)$. If in the momentum representation, there is not much difference between  the action of $D\SU(2)$ and $D\SO(3)$, in configuration space the difference is important since in the $\SU(2)$ case, the translation symmetry in implemented in a non-linear manner in configuration space.
\medskip

These different results points toward  new interesting ideas to develop.

\begin{itemize}
\item \textbf{GFT model for a 4d Euclidean quantum gravity}: In \cite{holo2}, a new spinfoam model for Euclidean quantum gravity was introduced using the spinor representation. It has the nice feature that the simplicity constraints are implemented through a Gupta-Bleuer procedure at the level of the spinors in a strong way \cite{holo1}. The construction we have presented here extends in a natural way to Ooguri's GFT on $\Spin(4)\sim\SU(2)\times\SU(2)$ which describes $BF$ theory in 4d. The next step would be to derive the GFT  for the spinfoam model presented in \cite{holo2} and understand how these spinfoam amplitudes can be written as the Feynman diagrams of a non-commutative field theory in the spinor variables.
    This is currently under development.

\item\textbf{ Loop quantum gravity as Voros non-commutative geometry}: Recently, non-commutative techniques were applied to  LQG to provide a non-commutative representation of the flux algebra \cite{fluxLQG}. The key idea  was to consider the  plane-wave used for $\SO(3)$. We can now reproduce this analysis using the spinor  representation together with the $\star$-product we have defined. As a consequence,  the flux algebra would be written as a Voros non-commutative algebra. The implications of this new representation should be explored.

\item\textbf{ Renormalization of  GFT}: The renormalization of the GFT's is a necessary step towards understanding the semi-classical regime and continuum limit of spinfoam models. A lot of work has been devoted to understand the renormalization features of Moyal non-commutative field theory. Since we have here rewritten the GFT as a Voros field theory, it would be interesting to see if the tools developed for Moyal can also be used in the spinfoam context \cite{renormalization-moyal}. As a first step, one could take advantage of the map relating the Voros and the Moyal $*$-products. It could be enlightening to see what is the meaning of this map in our context.

\item\textbf{ 4d Bicovariant differential calculus}: It is striking that  the 4d bicovariant differential calculus naturally emerges from the spinor representation of $\SU(2)$. It would be interesting to see if there exists (already?) a deeper mathematical structure which would explain this.

\item\textbf{Generalization of the $\star$-product to arbitrary Lie groups}: In the present paper, we have presented a group Fourier transform for the Lie group $\SU(2)$ on its Schwinger representation in terms of spinors (at the classical level). Introducing the spinorial plane-waves allowed us to define a $\star$-product dual to the convolution on $\SU(2)$, which actually matches the Voros product on the spinor variables. This procedure seems to be easily generalizable to more complicated semi-simple Lie groups that admit such a spinorial representation. We would then be able to define the $\star$-products dual to the convolution on these groups and relate them to the much simpler Voros product defined from the spinorial phase space structure.

\end{itemize}

\appendix

\section{$\SU(2)$ Coherent States in Term of Spinors} \label{CS}

Starting with a spinor $z\in\C^2$, for which we use a bra-ket notation:
$$
|z\ra=\mat{c}{z_0\\ z_1},\qquad
\la z|=\mat{cc}{\bz_0&\bz_1},
$$
with the canonical Poisson bracket $\{z_a,\bz_b\}=-i\delta_{ab}$, we quantize the components of $|z\ra$ and $\la z|$ respectively as annihilation and creation operators $a_{0,1},a^\dag_{0,1}$ acting on the Hilbert space $\cH_{HO}\otimes\cH_{HO}$ where $\cH_{HO}$ is the standard Hilbert space for a harmonic oscillator with basis $|n\ra$:
$$
a_0\,|n_0,n_1\ra_{HO}=\sqrt{n_0}\,|n_0-1,n_1\ra_{HO},\qquad
a_0^\dag\,|n_0,n_1\ra_{HO}=\sqrt{n_0+1}\,|n_0+1,n_1\ra_{HO}\,.
$$
Then quantizing the components of the 3-vectors $\vec{X}=\la z|\vsigma |z\ra$, we get the generators of the $\su(2)$ Lie algebra and its Casimir:
\be
J_3= \f12(a_0^\dag a_0-a_1^\dag a_1),\quad
J_+=a_0^\dag a_1,\quad
J_-=a_0 a_1^\dag=J_+^\dag\,, \quad
\hat{j}\,=\f12(a_0^\dag a_0+a_1^\dag a_1)
\ee
$$
[J_3,J_\pm]=\pm J_\pm,\qquad
[J_+,J_-]=2J_3,\qquad
[\hat{j},\vec{J}]=0\,.
$$
Diagonalizing the operators $\hat{j}$ and $J_3$ ,we recover the standard basis $|j,m\ra$ of $\su(2)$ irreducible representations and show that $\cH_{HO}\otimes\cH_{HO}=\bigoplus_{j\in\N/2}V^j$~:
\be
|j,m\ra=|n_0,n_1\ra_{HO},\qquad
\textrm{with}\quad
\left|
\begin{array}{l}
n_0=j+m\\
n_1=j-m
\end{array}
\right.\,.
\ee

Next, we introduce the coherent states for the harmonic oscillators:
$$
|z_0,z_1\ra_{HO}\,\equiv\,
\sum_{n_0,n_1} \f{z_0^{n_0}z_1^{n_1}}{\sqrt{(n_0)!(n_1)!}}\,|n_0,n_1\ra\,,
$$
from which we define the $\SU(2)$ coherent states by projecting them onto fixed values of the spin $j\in\N/2$:
\be
|j,z\ra
\,\equiv\,
\f{(z_0a_0^\dag+z_1a_1^\dag)^{2j}}{\sqrt{(2j)!}}\,|0\ra
\,=\,
\sum_{m=-j}^{+j} \sqrt{\f{(2j)!}{(j+m)!(j-m)!}}\,z_0^{j+m}z_1^{j-m}\,|j,m\ra\,,
\ee
$$
|z_0,z_1\ra_{HO}\,=\,\sum_j \f{1}{\sqrt{(2j)!}}\,|j,z\ra\,.
$$
These coherent states transform covariantly under the $\SU(2)$-action generated by the operators $\vec{J}$ (for more details, see e.g. \cite{un3,holo1}):
\be
e^{i\vec{u}\cdot\vec{J}} \,|j,z\ra =|j,e^{\f i2\vec{u}\cdot\vsigma}\,z\ra\,,
\ee
where $e^{\f i2\vec{u}\cdot\vsigma}$ is the representation for the group element $e^{i\vec{u}\cdot\vec{J}}$ in the fundamental two-dimensional representation of $\SU(2)$.
From this fundamental property of the $\SU(2)$ coherent states, it is straightforward to deduce that they are all obtained through the action of $\SU(2)$ group elements on the highest weight vector $|j,j\ra$ and that they are simply the tensorial powers of the coherent states in the fundamental $j=\f12$ representation:
$$
|j,z\ra= (\sqrt{\la z|z \ra})^{2j}\,g(z)\,|j,j\ra,\quad
g(z)=\f1{\sqrt{\la z|z \ra}} \mat{cc}{z_0 & -\bz_1 \\ z_1 & \bz_0}\quad
g(z)\,\mat{c}{1\\0}=\f1{\sqrt{\la z|z \ra}} \mat{c}{z_0\\z_1}\,,
$$
$$
|j,j\ra=|\f12,\f12\ra^{\otimes 2j},\qquad
|j,z\ra= |\f 12, z\ra^{\otimes 2j}= |z\ra^{\otimes 2j}\,.
$$
In particular, this allows to compute the matrix elements of $\SU(2)$ group elements on the coherent states:
\be
\la j,w|g|j,z\ra=\la w|g|z\ra^{2j}\,,\qquad
\la j,w|j,z\ra=\la w|z\ra^{2j}\,.
\ee
Moreover, we can write a decomposition of the identity on the Hilbert space $V^j$ of the irreducible representation of spin $j$ (for more details, see e.g. \cite{un3,holo1,holo2}):
\be
\id_j=\f1{(2j)!}
\int_{\C^2} \f{d^4z}{\pi^2}\,e^{-\la z|z\ra}\,
|j,z\ra\la j,z|\,.
\ee
In particular, this allows us to write the decomposition of the $\delta$-distribution on $\SU(2)$ onto characters as a Gaussian integral over the spinor variables:
\be
\delta(g)=
\sum_{j\in\N/2} (2j+1)\chi_j(g)
=
\f1{\pi^2}
\int d^4z\,e^{-\la z|z\ra}\,
\sum_j \f{(2j+1)}{(2j)!} \la j,z|g|j,z\ra
=
\f1{\pi^2}
\int d^4z\,e^{-\la z|z\ra}\,
(\la z|g|z\ra+1)\,
e^{\la z|g|z\ra}.
\ee
Working from there, we get by integration by parts:
$$
\delta(g)
=\f1{\pi^2}
\int d^4z\,e^{-\la z|z\ra}\,
(z\pp_z +1)\,
e^{\la z|g|z\ra}
=\f1{\pi^2}
\int d^4z\,(-\pp_z z +1)e^{-\la z|z\ra}\,
e^{\la z|g|z\ra}
=\f1{\pi^2}
\int d^4z\,(\la z|z\ra-1)e^{-\la z|z\ra}\,
e^{\la z|g|z\ra}.
$$

Following \cite{un3}, we can then define by group averaging the Livine-Speziale coherent intertwiners, which are $\SU(2)$-invariant states in the tensor product of $N$ irreducible representations:
\be
||\{j_i,z_i\}\ra\,\equiv\,
\int_{\SU(2)} dg\,
\otimes_i^N g\,|j_i,z_i\ra
\quad
\in\textrm{Inv}[\bigotimes_i^N V^{j_i}]\,.
\ee
Finally, by summing over the spin labels, we define the coherent intertwiner states, which diagonalize the intertwiner annihilation operators and which are labeled only by spinor variables as introduced in \cite{holo1,holo2}:
\be
|\{z_i\}\ra\,\equiv\,
\sum_{\{j_i\}}
\prod_i\f{1}{\sqrt{(2j_i)!}}
||\{j_i,z_i\}\ra
\,=\,
\int_{\SU(2)} dg\,
\otimes_i^N g\,|z_i\ra_{OH}\,.
\ee
These coherent intertwiners are covariant under the action of $\U(N)$ as shown in \cite{un3,holo1,holo2}. We can compute their norms and scalar products either by using the $\U(N)$ structure or by computing directly the integrals over $\SU(2)$.

\section{On the choice of plane-wave}
\label{kernel1}

An interesting choice for the plane-wave, motivated from \cite{holo2}, is to consider the plane-wave and measure
\bes
\cE_g(z)= e^{-\la z|z\ra}\,
\frac{(\la z|g|z\ra+1)}{\la z|z\ra +1}\,
e^{\la z|g|z\ra},\qquad
d\mu(z)= d^4z (\la z|z\ra +1),
\ees
with the $\star$-product, as usual reflecting the $\SU(2)$ group structure and relevant measure
\bes
( \cE_{g_1}\star \cE_{g21}) (z) &=&  \cE_{g_1g_2}(z) \equiv e^{-\la z|z\ra}\,
\frac{(\la z|g_1g_2|z\ra+1)}{\la z|z\ra +1}\,
e^{\la z|g_1g_2|z\ra} \nn
\eeq
It is normalized since $\cE_{1}(z)=\one$. This plane-wave and $\star$-product were suggested in \cite{holo2} to construct the partition function for the $BF$ theory. In this case, the delta function in the configuration space is given by
\beq
\delta_{\star_{bf}}(\vec X)= 2 \frac{e^{-\ka |\vec X|}}{1+ \kappa |\vec X|}
\eeq
Different plane-waves lead therefore to different realizations of the delta function on configuration space.

\medskip

In a general manner, we can construct a plane-wave $\cK_g(z)$  and introduce a measure $d\mu(z)$ such that
\bes
\left.
\begin{array}{c}
\cK_g(z)=\sum_j \f{\alpha_j}{(2j)!}\la z|g|z\ra^{2j} \\ d\mu(z)= \sum_j \f{\beta_j}{(2j)!}\la z|z\ra^{2j}
\end{array}
\right\}
\quad
\Rightarrow
\quad
\delta(g)= \int d\mu(z)\, \cK_g(z),
\ees
which puts constraints\footnotemark{ } on the coefficients $\alpha_i$ and $\beta_i$.
\footnotetext{We can further demand it to be an element of $L^2(\SU(2))$, which translates into
$$
\sum_j \f{|\alpha_j|^2}{(2j)!^2} (\la z|z\ra^2)^{2j}\,<\,+\infty,\quad\forall z
$$
}
The $\star$-product is  defined as usual to reflect the group product.
\be
\cK_{g_1}\star\cK_{g_2}(z)=\cK_{g_1g_2}(z)
\ee
The plane-wave $\cK_g(z)$ is in general not normalized, in the sense that
$
\cK_1(z)\neq \one.
$
We can demand to normalize it, in which case we consider the renormarlized plane-wave, new $\star$-product and new measure
\beq
\tilde \cK_g(z)=  \cK^{-1}_1 (z)\cK_g(z), \quad (\tilde \cK_{g_1}\star \tilde \cK_{g_2})(z)=\tilde \cK_{g_1g_2}(z) , \quad d\mu(z)\dr d\mu(z) \cK_1(z)= d\tilde \mu(z).
\eeq
%
Finally, we can demand as well that the coordinates $X_i\one_\star$ are given in terms of the derivative of the plane-wave evaluated at the identity.
\beq
 && -i  \int dg\, \delta(g) \frac{\partial}{\partial p^i} \cK_g(z)=  X_i \one_\star\,  \Leftrightarrow  X_i \sum_{k\in \N^*/2} \frac{\alpha_k }{(2k-1)!} (\ka |\vec X|)^{2k-1}  =    X_i  \cK_1(\vec X) . \label{X}
\eeq
This  leads to a plane-wave $\cK^*_g(z)$,  $\star$-product and measure
\beq
\cK^*_g(z)= \cX(g,z) e^{\la z |g|z\ra},\quad (\cK^*_{g_1}\star\cK^*_{g_2})(z)= \cK^*_{g_1g_2}(z), \quad d\mu(z)=\frac{\la z|z\ra -1}{\cX(z)} e^{- \la z|z\ra},
\eeq
where $\cX(g,z)$ is function invariant under $z\dr e^{i\alpha}z$ (that is in $\cC_\star(\R^3)$ ) which is zero nowhere and which satisfies $\frac{\partial\cX(g,z)}{\partial p}_{|_{p=0}}=0$ . As said before we can then normalize $\cK^*_g(z)$ so that $\one_\star=\one$, in which case, $\cK^*_g(z)= e^{-\la z|z\ra}e^{\la z|g|z\ra}= \tilde  E_g(z)$ if $\cX(g,z)=1$ and $\cK^*_g(z)= e^{i\vec X\cdot \vec p}$ if $\cX(g,z)= e^{-|\vec X|\tr g}$. Note however that the latter case does not describe the full structure of $\SU(2)$ as recalled in section \ref{so3}.

\smallskip

As a conclusion of this discussion, we see that the nicest plane-waves relevant for $\SU(2)$ are with no surprise the exponential type $\cK^*_g(z)= e^{\la z|g|z\ra}= E_g(z)$ or $\cK^*_g(z)= \tilde  E_g(z)=e^{-\la z|z\ra}e^{\la z|g|z\ra}$  which we have considered earlier.


\end{document}